\documentclass[aps,fleqn,nofootinbib,onecolumn,groupedaddress]{revtex4}

\usepackage[dvipdfmx]{graphicx}

\usepackage[T1]{fontenc}
\usepackage{ae,aecompl}
\usepackage{grffile}
\usepackage{enumitem}
\usepackage{gensymb}

\def\be{\begin{equation}}
\def\ee{\end{equation}}
\def\ba{\begin{eqnarray}}
\def\ea{\end{eqnarray}}

\newcommand\jcap{JCAP}
\newcommand\mnras{Mon.~Not.~Roy.~Astron.~Soc.}

\newcommand\araa{Ann.~Rev.~Astron.~Astrophys.}
\newcommand\apjl{ Astrophys.~J.,~Lett.}
\newcommand\apjs{Astrophys.~J.,~Suppl.}
\newcommand\apss{ Astrophysics and Space Science}
\newcommand\aap{  Astron.~Astrophys.}
\newcommand\aapr{ Astron.~Astrophys.~Rev.}
\newcommand\aaps{ Astron.~Astrophys.,~Suppl.}
\newcommand\procspie{Proc.~SPIE}
\newcommand\ssr{Space~Sci.~Rev.}
\newcommand\physrep{Phys.~Rep.}
\newcommand\nar{New~Astron.~Rev.}

\newcommand{\clvvpop}{C_{\ell}^{\rm VV, \rm Pop III}}
\newcommand{\clvvg}{C_{\ell}^{\rm VV, \rm MW}}
\newcommand{\clvveg}{C_{\ell}^{\rm VV, \rm EG}}
\newcommand{\clvv}{C_{\ell}^{\rm VV, \rm MW}}
\newcommand{\clii}{C_{\ell}^{II,\rm MW}}
\newcommand{\clvvs}{C_{\ell}^{\rm VV,\rm SOI}}
\newcommand{\clvvprim}{C_{\ell}^{\rm VV,\rm prim}}
\newcommand{\clvvgc}{C_{\ell}^{\rm VV,\rm galaxy\ \rm cluster}}
\newcommand{\clnoise}{I_{\ell}^{\rm VV}}
\newcommand{\hammu}{\texttt{HAMMURABI}}
\newcommand{\msun}{M_{\odot}}
\usepackage{graphicx}	
\usepackage{amsmath}	
\usepackage{amssymb}	
\usepackage{hyperref}
\usepackage{bm}
\usepackage{color}

\hypersetup{
    colorlinks=true,
    linkcolor=red,
    citecolor=blue,
}
\begin{document}
\title{Circular polarization of the CMB: Foregrounds and detection
prospects}

\author{Soma King$^{1}$, Philip Lubin$^{2}$}

\affiliation{
$^1$Department of Physics and Astronomy, University of California Davis,
Davis, CA 95616 \\
$^2$Department of Physics, UC Santa Barbara, Santa Barbara, CA  93106
}




\begin{abstract}
The cosmic microwave background (CMB) is one of the finest probes of
cosmology. Its all-sky temperature and linear polarization (LP) 
fluctuations have
been measured precisely at a level of $\delta T/T_{\rm CMB}$ $\sim$
10$^{-6}$. In comparison, circular polarization (CP) of the CMB, however, has not been precisely
explored. Current upper limit on the CP of the CMB is at a level of
$\delta V/T_{\rm CMB}$ $\sim$ 10$^{-4}$ and is limited on
large scales. Some of the cosmologically important 
sources which can induce a CP in the CMB include early universe symmetry
breaking, primordial magnetic field, galaxy clusters and Pop III
stars (also known as the First stars). Among these sources, Pop III stars
are expected to induce the strongest signal with levels strongly
dependent on the frequency of observation and on the number, N$_{\rm
p}$, of the Pop III stars per halo. Optimistically, a CP signal in the CMB due
to the Pop III stars could be at a level of $\delta V/T_{\rm CMB} \sim$ 2 $\times$
10$^{-7}$  in scales of 1$\degree$ at 10 GHz, which is much smaller than
the currently existing upper limits on the CP measurements.
Primary foregrounds in the cosmological CP detection will come from the galactic
synchrotron emission (GSE), which is naturally (intrinsically) circularly
polarized. We use data-driven models of 
the galactic magnetic field (GMF), thermal electron density and relativistic
electron density to simulate all-sky maps of the galactic CP.
This work also points out that the galactic CP levels are important below 50
GHz and is an important factor for telescopes aiming to detect
primordial B-modes using CP as a systematics rejection channel.
In this paper, we
focus on S/N evaluation for the detectability of the Pop III induced CP signal in the
CMB. We find that
a S/N higher than unity is achievable, for example, with a 10 m
telescope and an
observation time of 20 months at 10 GHz, if N$_{\rm p}\geq$ 100. We also
find that, if frequency of observation and resolution of the beam is appropriately
chosen, a S/N higher than unity is possible with N$_{\rm p}$ $\geq$ 10
and resolution-per-pixel $\sim$ 1 $\mu$K at an observation time of 60
months. A summary of different sources which can induce a CP in the CMB
is summarized in Table.~\ref{table:table1}.  Final results related to
the detectability of cosmologically important CP are summarized in
Fig.~(\ref{fig:snr_grid1}-\ref{fig:snr_grid1_dg}). 
\end{abstract}
\maketitle

\section{Introduction}
\label{sec:intro}
The CMB has been a finest probe
of cosmology. A complete characterization of the CMB is made by
quantifying the four Stokes parameters \citep{jones} associated to its
unpolarized intensity, LP and CP. 
Temperature and polarization anisotropies of the CMB 
have provided invaluable
insight into the universe that we live in. Venture into the CMB studies
began with its discovery in 1965 \citep{penzias, dicke} which showed the 
CMB radiation to be a Black-body with a mean temperature of T $\sim$ 2.7255 K.
Subsequently, the first detection of angular variation of the CMB mean 
temperature was measured with the COBE satellite \citep{cobe} and today,
the temperature fluctuations have been measured with an unprecedented
precision \citep{dasi,wmap, planckgeneral, act-temp,spt-temp}.
CMB is also linearly polarized due to Thomson scattering at the surface
of last scattering. This polarization level is at $\sim$ 5$\%$ of the
temperature anisotropies. Polarization fluctuations have been
measured with a high precision and many future experiments are aimed to
increase the precision of polarization measurements \citep{bicep2-lensing, polarbear, quiet,spt-lensing,
act-lensing}. The Planck experiment reached a 
sensitivity level, $\delta T/T_{\rm CMB}$ $\sim$ 10$^{-6}$
for temperature and polarization, a much higher level of sensitivity
compared to the WMAP satellite.
Together, the
CMB temperature and LP data bring
an overwhelming support for the cosmological standard model or the
$\Lambda$CDM
\citep{bond} model. It is only natural to wonder what
information does the CP of the CMB store? The CP of the CMB has not been
explored extensively and the current level of CMB CP measurement stands
at a level $\Delta V/T_{\rm CMB}$ $\sim$ 10$^{-4}$ in the scales of 8$\degree$ and
24$\degree$ \citep{mainini}. This upper limit is much higher than the
level of CP signal expected from any cosmologically relevant sources.
Previously, \cite{lubin} had made one of the first efforts to measure to
the CP of the CMB. Recently, CLASS \citep{class} and PIPER \citep{piper}
experiments proposed to
measure the LP and CP of the CMB, however, the experiment is designed
to focus on primarily the measurements of the LP of the CMB. These
experiments aim to detect primordial B-modes using variable-delay
polarization (VPM) instruments where
the observing strategy relies on the expectation of 
CP of the sky to be null to constrain systematic uncertainties. In this
paper, we have provided galactic CP maps for frequencies relevant for
both CLASS at 40 GHz and PIPER at 220 GHz, showing that the galactic 
CP effects are important for frequencies below 50 GHz.

There are various cosmologically important
sources which may induce a CP in the CMB via different
mechanisms. Some of these CP production channels are intrinsic to the
emission from a certain type of source, for example synchrotron emitting
radio sources \citep{sazonov, westfold}. Some are due to the effects of external
magnetic fields \citep{ensslin} or other birefringent effects, and
finally some mechanisms propose CP generation in the CMB by
models that stand on departures from the Standard model of particle
physics \citep{thomson,compton, noncomm, noncomm-1, cnub,lv}.

Among the sources which can induce a CP in the CMB due to presence
of a magnetic field, are the so called  Pop III stars, also known as the
First stars. See (\cite{bromm, pop3-properties}) for a review of the Pop III
stars. \cite{decp} describes how a CP in the CMB can be induced by
the remnants of Pop III stars that went supernovae (SNe). These 
stars, residing in dark matter mini-halos, provide a window into the early structure formation which ended the
cosmological dark ages and began the re-ionization along with metal
enrichment of the inter-galactic medium. Pop III stars are expected to
exist based on the numerical simulations of primordial stars formation
and fossil abundance of SNe. However, there are no definite 
constraints on the
properties of these stars \citep{brommpop3openqs,currentprobpop3}. These
stars are generally $not$ expected to be directly detected by
the most advanced future space telescopes like the WFIRST and the JWST \citep{jwst}, except under
certain conditions when these Pop III stars explode into pair instability
SNe \citep{pinspop3} or hyper nova \citep{hypernova} releasing energies
around $\sim$ 10$^{53}$ ergs.
These stars and their properties are speculated by numerical simulations
\citep{tomsim,ragesim}, however, they are far from being verified
by observations. CP of the CMB provides an
indirect and a much economical way of exploring into these Pop III
stars.

In this paper we will primarily
focus on S/N determination of the CP signal in the
CMB due to the Pop III stars. However, there are other cosmologically
important sources which induce CP in the CMB. These sources include the
primordial magnetic field (PMF) \citep{thomson, compton}, different modifications and symmetry
breaking mechanisms \citep{noncomm, noncomm-1, cnub} beyond the Standard model particle physics and the galaxy clusters
\citep{cooray}. Most of these sources induce a lower signal level in CP
of the
CMB when compared to the level of CP induced by the Pop III stars.
However, these sources could certainly be explored via the CP in the CMB
once the instrumental sensitivity improves.

In addition to the cosmologically important sources of CP, 
the Milky Way (MW) galaxy produces
synchrotron radiation which is intrinsically circularly polarized.
Circularly polarized synchrotron emission from the MW galaxy acts as a
foreground towards the detection of the cosmological CP in the CMB.
Currently, there is not enough observational data to accurately 
shed light on
the level of CP from the GSE. 
In this paper,
we generate numerical simulations of the galactic CP due to synchrotron
emission using data-driven models of the GMF and cosmic ray electron energy
distribution.

CP in the CMB could potentially detect the existence of the Pop III
stars, symmetry breaking in the early universe or the existence of
primordial magnetic field. Implementing direct detection of these
sources will need a revolution on the instrumentation front, involve a long
time-scale and a very high cost. Exploring some of 
these highly interesting
sources indirectly via the CP in the CMB is possible within the current
reach of instrumentation, achievable at a moderate timeline and
cost.

In this paper, we will discuss the sources, foregrounds and detection
prospects of the cosmological CP of the CMB. Sec.~\ref{sec:transfer} discusses theoretical
framework needed for the description of CP, Sec.~\ref{sec:overview} presents an overview
of various sources and mechanisms which induce CP in the CMB. Sec.~\ref{sec:foregrounds}-\ref{sec:maps}
discuss the galactic foregrounds
in CP and simulations. Sec.~\ref{sec:detection} represent the results on detection
prospects. Finally, we discuss the future directions and implications of
this work in Sec.~\ref{sec:discussion}.
\section{Polarization transfer equation}
\label{sec:transfer}
\begin{figure}
\begin{center}
\includegraphics[height=0.5\textwidth]{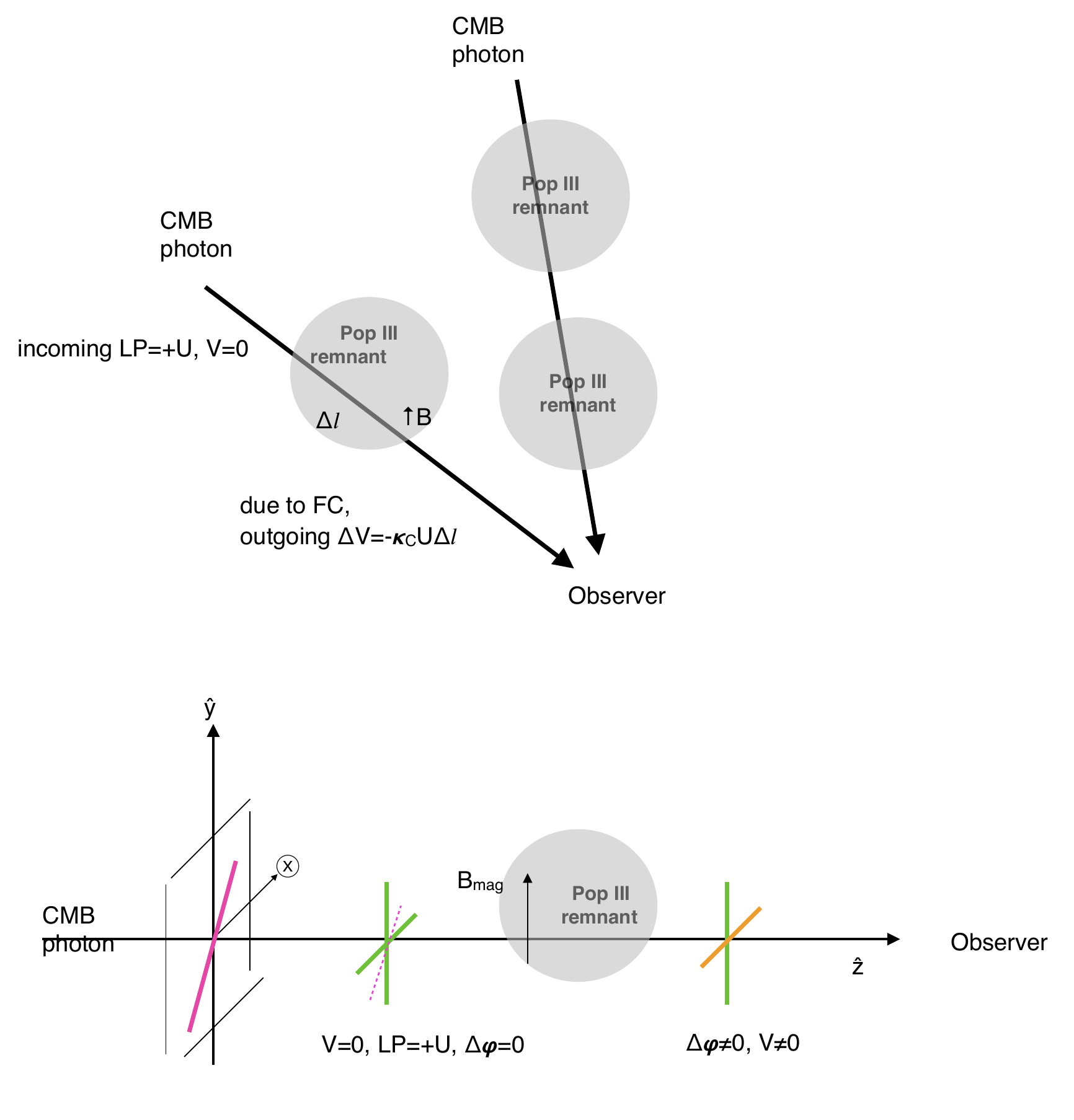}
\end{center}
\caption{
Schematic view of the generation of CP via the FC: Top frame shows the overall geometry of the 
problem. Bottom frame shows how exactly the FC mechanism generates
CP when linearly polarized CMB passes through the
magnetized plasma of the Pop III remnants. The
phase difference, $\Delta \phi$, between the mutually orthogonal E-field
components is zero and represents a only a linearly polarized CMB prior
to its passage through the Pop III remnant. The magnetic field in the
remnant creates an asymmetry due to the 
difference in Lorentz forces between
the charged particles moving along $\hat{y}$ and $\hat{x}$. This now results in
a non-zero phase difference between the mutually orthogonal E-field
components of
the outgoing wave and hence creates a circularly polarized wave. 
}
\label{fig:qu}
\end{figure}

A complete description of polarization of an electro-magnetic (EM) wave is
described by four Stokes parameters, intensity I, linear
polarizations (LP) Q and U, and circular
polarization (CP) V. Together (I,Q,U,V) represent the Stokes vector
associated to the EM wave. The evolution of different Stokes
vector components of an EM wave propagating through a plasma are 
governed by the following polarization transfer equation \citep{jones}.
\begin{equation}
\left(\begin{array}{c} dI/dz \\ dQ/dz
\\dU/dz\\ dV/dz \end{array} \right)=
\left(\begin{array}{c} \eta_I \\ \eta_Q \\\eta_U \\\eta_V \end{array} \right)
+\left(\begin{array}{cccc}-\kappa_I & -\kappa_Q & -\kappa_U &-\kappa_V \\-\kappa_Q &-\kappa_I& -\kappa_F&
-h_Q \\-\kappa_U &\kappa_F
& -\kappa_I & -\kappa_C \\ -\kappa_V &h_Q & \kappa_C &-\kappa_I \end{array} \right) \left( \begin{array}{c} I \\Q \\U\\V \end{array}\right)
\label{eq:transfer}
\end{equation}
where the spatial derivatives on the left side of the equation indicate
the change
in Stokes vector along the line of sight, taken to be inclined along
the z axis. The coefficients $\eta_{I,
Q,U,V}$ indicate emissivity and $\kappa_{I,Q,U,V}$ indicate absorption
coefficients corresponding to the Stokes vectors I, Q, U and V. Under an
isotropic distribution of unperturbed particles in a plasma, the conversion
coefficient, $h_{\rm Q}$ between Q and V, vanishes due to dielectric
symmetries \citep{jones,beckert}. Eq.~(\ref{eq:transfer}) uses a coordinate
frame where the sky projected magnetic field component is aligned along
the y axis. Under this geometry, Stokes components +U is defined as
LP aligned along an axis which makes an angle of 
45$\degree$, clockwise with respect to the y axis. Component +Q(-Q) is
defined as LP aligned along the y(x) axis.
Please see
Fig.~\ref{fig:qu} for a schematic view of the coordinate system used.
Stokes components I and V are invariants under coordinate
transformations \citep{kosowsky}.

The ordinary rotation between Stokes components Q and U are
driven by the Faraday rotation (FR) coefficient, $\kappa_F$. FR is
simply the rotation of the plane of LP of an EM wave
propagating through a magneto-active plasma due to a local 
magnetic field along the line of sight. 

The generalized rotation,
also known as the Faraday conversion (FC), described by the conversion
coefficient, $\kappa_C$, controls the transfer between U and V
components. To understand the FC effect, consider a linearly polarized
EM wave propagating through a plasma where an external magnetic field is aligned along the y
axis. Charged particles moving along the y axis will experience a different
Lorentz force than those along the x axis. Let the LP
of the incoming EM wave be +U (for example), which can be decomposed into two
linear components, each along the y and the x axes. Since the particles
along the x and y axes move differently due to the asymmetry introduced
by the external magnetic field, there will now be a phase difference between
the orthogonal LP components (along x and y axes). Thus EM wave will
therefore have acquired a CP, depicted by the Stokes component V. 

In the case of the CMB, under the standard cosmological model, $\eta_V$
vanishes \citep{weinbergbook}. 
The coefficients $\eta_{Q, U}$ are non-zero due to
anisotropic Thomson scattering.  Effects of
non-standard cosmology or new physics can induce a  non-zero $\eta_V$.
We provide a summary of such possibilities in
Sec.~\ref{sec:summary} and in Table.~(\ref{table:table1}). 
Conversion between U and Q components of the 
CMB due to FR induced by the galactic magnetic field has been 
been studied by \cite{defr}.

Possibilities of conversion between U and V components  of the CMB 
due to
the FC effects induced by the Pop III remnants and galaxy clusters are 
discussed in Sec.~\ref{sec:pop3} and
Sec.~\ref{sec:gc}, respectively.

Eq.~(\ref{eq:transfer}) also applies to the GSE. Under a coordinate system where the 
sky-projected magnetic field is inclined
along the y-axis, $\eta_U$ and $\kappa_U$ of the synchrotron emission 
are set to be zero. For the GSE, 
$h_Q$ is also set to be zero \citep{beckert}. 

A detailed description of the generalized transfer
equation in the context of synchrotron emission 
can be found in \cite{sazonov,beckert}. In this paper we adopt
the symbols and conventions from \cite{beckert} to describe the polarization
relevant quantities.   
\section{Overview of different CP sources}
\label{sec:overview}
\subsection{Pop III stars as CP sources in the CMB}
\label{sec:pop3}
Pop III stars mark the transition of a simple homogeneous universe of H-He
gas into a complex and structured universe after the cosmic dark ages.
Pop III stars are therefore also known as the First stars. Simple
interpretation of optical 
depth observed by WMAP and PLANCK suggest star formation activity at z
$\ge$ 11 \citep{plancktau,yoshida,dunlop}. Supernovae explosions of the Pop III
stars was responsible for the metal enrichment of the inter-galactic
medium \citep{loeb}. These Pop III stars are predicted to form
in dark matter minihalos of mass $\sim$ 10$^{6-7}\msun$ around z $\sim$
20-30 \citep{brommpop3openqs}. A low-mass halo is needed to virial
temperature below the threshold of T$_{\rm vir}$=10$^4$K to allow
efficient atomic hydrogen cooling, necessary for collapse. The Pop III
stars are yet unobserved and there is significant lack of certainty of
their properties, most of which are predicted from numerical simulations
and therefore, are highly model dependent.

Significance of understanding of the Pop III stars is enormous. 
Many of the implications
drawn from the Pop III stars depend on the mass of these primordial stars. If the Pop III stars are
massive ($\geq$ 100 $\msun$), they can be connected to several effects
which can be tested in the distant future. For example, Sunyaev Zeldovich
effects of the CMB \citep{haiman,oh}, Gravity waves from Black
holes formed from Pop III remnants \citep{madau,suwa}. One of the most
certain and significant effects of the Pop III stars is the cosmological
heavy element production and the cosmic reionization \citep{frebel,
tumlinson,greif, wyithe}.

In \cite{decp}, Pop III stars are established as a source of
appreciable CP in the CMB. 
Intrinsic CP of such sources is small, however, the
CMB acts as a back-light in this scenario. 
As the CMB photons pass through the
relativistic plasma of the Pop III remnants, a fraction of the 
CMB linear
polarization is converted into CP, via the FC mechanism. Please
refer to Sec.~\ref{sec:transfer} for a schematic explanation of the FC
mechanism under which $\kappa_C$ in
Eq.~(\ref{eq:transfer}) describes
the transfer of the Stokes U into the Stokes V component in the CMB.

Using a simple analytical model of a SN
remnant of a Pop III star,
\cite{decp} evaluates $\kappa_C$ described in
Eq.~(\ref{eq:transfer}) as
\begin{equation}
 \kappa_C \sim 20~{\rm pc}^{-1}~\left(
\frac{t_{\rm age}}{10^6 {\rm yr}} \right)^{-\frac{12}{5}} \left( \frac{E_{\rm SN} }{10^{53} {\rm
ergs}} \right)^{\frac{4}{5}}
\left( \frac{1+z}{20} \right)^{\frac{3}{5}} \left( \frac{f_{\rm mag}}{0.1}\right)
\left( \frac{f_{\rm rel}}{0.1}\right)
\left( \frac{\nu}{1{\rm GHz}}\right)^{-3}
\label{eq:kc}
\end{equation}
where $t_{\rm age}$ is the age of the SN remnant, $E_{\rm SN}$ is the
energy of the explosion and
$f_{\rm mag}$ and $f_{\rm rel}$ are the fractions of the explosion
energy respectively into the relativistic electron energy and the
magnetic fields in a SN remnant. $\nu$ is the CMB
observation frequency. Following the so-called halo model, the angular power spectrum of the
CP due to the Pop III stars is evaluated. Angular power spectrum is
effectively the square of the rms fluctuation in V, or 
$\ell(\ell+1)C_\ell^{VV}/(2\pi) \sim (\delta V)^2$. In \cite{decp}, Pop
III stars are assumed to only exist in halos with virial temperature,
$T_{\rm vir}>10^4$ K where atomic hydrogen cooling is effective for the
collapse. Since the signal of CP due to FC mechanism falls off with
frequency, we set our normalization frequency to be 10 GHz in future
equations in this paper. A frequency much lower than 10 GHz calls for a
full solution of the transfer equation, which will be addressed in
future work. 

Around $\ell \sim 100$, a simple formula corresponding to the brightness
temperature associated to the fluctuation  $\delta$V can be expressed as the following.
\begin{eqnarray}
\delta V_{\rm Pop III}(\nu)|_{t_{\rm age}=10^4~yr,~\ell \sim 100)} &\sim 7
\times 10^{-2} \left(\frac{\nu}{10 \rm GHz}
\right)^{-3}\left(\frac{N_{\rm p}}{100}\right)\left(\frac{E_{\rm
SN}^{(16+2p_{\rm Pop III})/20}}{10^{53}\rm ergs} \right)~\rm \mu \rm K \nonumber \\
\delta V_{\rm Pop III}(\nu)|_{t_{\rm age}=10^4~yr,~\ell \sim 1000)} &\sim 8\times
10^{-1} \left(\frac{\nu}{10 \rm GHz}
\right)^{-3}\left(\frac{N_{\rm p}}{100}\right)\left(\frac{E_{\rm
SN}^{(16+2p_{\rm Pop III})/20}}{10^{53}\rm ergs}\right)~\rm \mu \rm K 
\label{eq:pop3}
\end{eqnarray}
where N$_{\rm p}$ is the number of Pop III stars per halo. $p_{\rm Pop
III}$ is the spectral index of the electron energy distribution around
the Pop III remnant, and $p_{\rm Pop III}$ $\sim$ 2
\citep{syncpop3}.

\begin{figure}
\begin{center}
\includegraphics[height=0.45\textwidth]{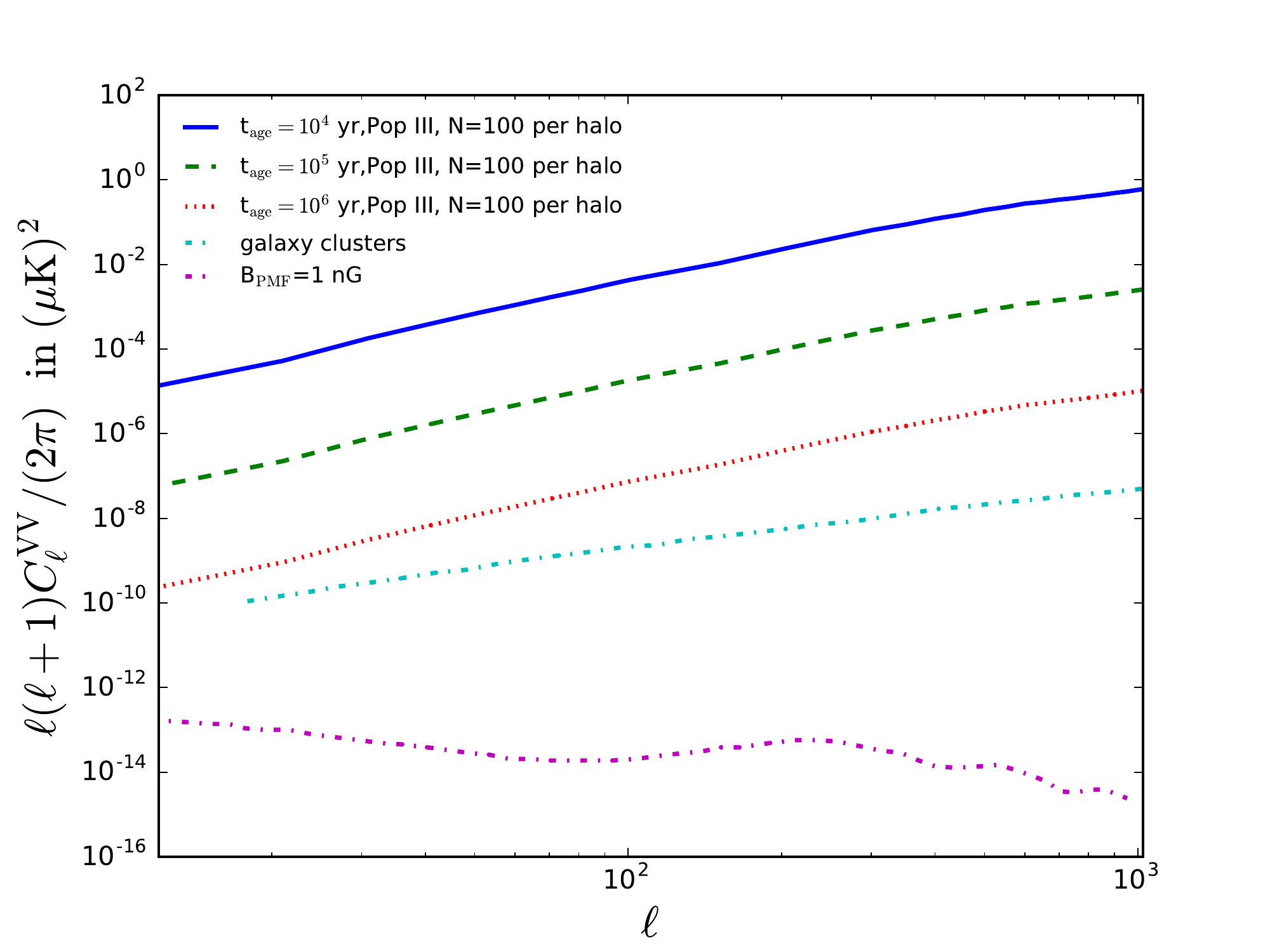}
\end{center}
\caption{
Angular power spectra of circular polarization in the CMB 
due to the Pop
III stars, galaxy clusters and primordial magnetic field  in ($\mu$K)$^2$. 
Pop III signal is produced via Faraday
conversion. The signal level shown depend on the number of Pop III stars per halo, age of the remnant, energy generated at collapse
and very strongly on the frequency of the CMB observation \citep{decp}. A 
few cases for the age of the remnant of the Pop III explosion, with
$10^4<t_{\rm age}<10^6$ in years are shown. 
The frequency of observation of the
CMB has been chosen to be $\nu=$ 10 GHz and the number of
Pop III stars is assumed to be 100 per halo. Also shown is the galaxy
cluster signal due to FC at $\nu$=10 GHz (taken from \citep{cooray}) and the signal due to
primordial magnetic field at $\nu$=10 GHz with $B_{\rm PMF}$=1 nG (taken
from \cite{thomson}). 
}
\label{fig:signal_all}
\end{figure}

In Fig.~\ref{fig:signal_all} different Pop III associated CP
signals are shown. Note that, $\clvvpop$ falls off very sharply with frequency of
CMB observation as
$\nu^{-6}$ and also with increasing age, $t_{\rm age}$ of
the Pop III remnants. In Fig.~\ref{fig:signal_all} we use
$10^4<t_{\rm age}(yr)<10^6$ which comes from the Compton cooling
timescale of the remnants. The number of Pop III stars in a halo is
uncertain and could be
up to $\sim$ 10$^3$ \citep{pop3-xu}. $\clvvpop$ increases with N$_{\rm
p}$ as N$_{\rm p}^2$. 
Not explicit from Eq.~(\ref{eq:pop3}), the quadratic 
dependence of the signal
in $\delta$V on
the local magnetic field, B is B$^2$. This dependence can be easily seen
in Eq.~(\ref{eq:kc}) via linear dependence of $\kappa_C$ on
$f_{\rm mag}$ or effectively the magnetic energy ($\sim B^2$).

All relevant cosmological and astrophysical 
parameters are also described in \cite{decp}. 

\begin{table}
\caption{Summary of various circular polarization (CP) sources with
levels showed in fluctuations of Stokes V
in the units of temperature (K). Shown also the dependence of the
fluctuation, $\delta$V
(K) on magnetic field (B) and frequency of the CMB
observation ($\nu$). A temperature corresponding to $\delta$V 
is derived by using
$\delta$V$\sim$ $\sqrt{\ell(\ell+1)C_{\ell}^{\rm VV}/(2\pi)}$ in K.
$\alpha_{\rm sync}$
is the synchrotron spectral index and for the galaxy $\alpha_{\rm sync}$
$\sim$ 2.8. $\ell$ gives a measure of the angular scale. t$_{\rm age}$
represents the age of the Pop III star remnants.}
\vspace*{1.0cm}
\centering
\resizebox{\textwidth}{!}{
\begin{tabular}{c c c c c c}
\hline\hline
Source & Mechanism for CP       & Frequency & B &Predicted CP \\ 
       &                        & dependence & dependence & signal in $\delta$V (K) \\
       &                        &            &            & at $\nu$=10 GHz\\[0.5ex]
\hline\hline
Primordial &primordial B+                      &$\nu^{-3}$ &B
&10$^{-9}$ \\
           &Compton scattering \citep{compton} &           &           &
\\
\hline
Primordial & Lorentz invariance&$\nu^{-3}$ & NA &10$^{-12}$\\
& violations \citep{lv} &  &  & &\\
\hline
primordial & Non-commutivity \citep{noncomm,noncomm-1} &$\nu^{-1}$ &NA
&10$^{-12}$\\
\hline
primordial &B+Thomson &
$\nu^{-3}$ & B$^2$ & 10$^{-12}$\\
 &scattering \citep{thomson} &  & &  & \\
\hline
Cosmic neutrino & Scattering with & $\nu^{-1}$ & NA &10$^{-8}$ \\
background& left handed           &            &    & \\
(C$\nu$B) & neutrinos \citep{cnub} &  & & \\
\hline
Pop III stars &FC \citep{sazonov,decp} & $\nu^{-3}$ & B$^{2}$&
few $\times$ 10$^{-6}$\\
 & &  & &($\ell$ $\sim$ 1000, $t_{\rm age}$=10$^{4}$ yr, $N_{\rm
p}$=100)\\
 & & & & few $\times$ 10$^{-5}$\\
 & & & & ($\ell$ $\sim$ 1000, t$_{\rm age}$=10$^4$ yr, $N_{\rm p}$=1000)\\
 & & & & few $\times$ 10$^{-7}$\\
 & & & & ($\ell$ $\sim$ 100, t$_{\rm age}$=10$^4$ yr, $N_{\rm p}$=100)\\
\hline
Galaxy clusters  &FC  & $\nu^{-3}$ & $B^2$ &10$^{-10}$\\
                 &    &            &       &($\ell$ $\sim$
1000 \citep{cooray})\\
\hline
Galactic synchrotron &intrinsic &$\nu^{\left(-2-\alpha_{\rm
sync}/2\right )}$ &B$^{3/2}$& 10$^{-8}$ ($\ell$ $\sim$ 100)\\
&emission \citep{westfold} &  & & $<$10$^{-9}$ ($\ell$ $\sim$ 500)\\[1ex]
\hline
\end{tabular}}
\label{table:table1}
\end{table}

\subsection{Galaxy clusters as a source of CP in the CMB}
\label{sec:gc}
Another source of CP in the CMB are the galaxy
clusters which induce CP via the FC mechanism due to their
magnetic field and relativistic electrons. 
This scenario was explored in \cite{cooray} where a mean field of 10
$\mu$G with a coherent scale of 1 Mpc was adopted. 
In Fig.~\ref{fig:signal_all} we also represent the $\clvv$ induced in
the CMB due
to FC in the galaxy clusters at $\nu=10$ GHz, a result which has been
derived from \cite{cooray}. FC coefficient in this scenario is given by,
\begin{equation}
k^{\rm galaxy\ cluster}_C \sim few \times 10^{-3}\left(\frac{\rm
Mpc}{\rm RM}\right)
\end{equation}
where RM is the rotation measure for the galaxy clusters. Temperature
equivalent Stokes V in the CMB due to galaxy clusters around $\ell \sim
1000$ can be simply expressed as
\begin{equation}
\delta V_{\rm galaxy\ cluster} \sim 3 \times 10^{-3}\left(\frac{\nu}{10 \rm
GHz}\right)^{-3} \mu K
\label{eq:gc}
\end{equation}
 
Note that the CP signal due to galaxy clusters is much smaller compared
to that due to the Pop III stars.  
\subsection{Primordial magnetic field as a source of CP in the CMB}
\label{sec:prim}
Primordial magnetic field (PMF) is speculated as a source of magnetic
field in the universe \citep{reviewpmf,ruth}. Current limits on the PMF
is $\sim$ few nG today \citep{planckpmf}. \cite{thomson} showed that 
CP is naturally generated in the CMB in
presence of Thomson scattering, primordial magnetic field,
spatial curvature and adiabatic fluctuations. This CP generation is
intrinsic ($\eta_V$) and does not require  any pre-existing linear
polarization, unlike FC. In Fig.~\ref{fig:signal_all} $\clvv$ due to
the presence of a primordial magnetic field of an equivalent field
strength, B$_{\rm PMF}$=1 nG is shown, following \citep{thomson}.
\begin{equation}
\delta V_{\rm prim} \sim 6 \times 10^{-7}\left(\frac{\nu}{10\rm
GHz}\right)^{-3}\mu K
\label{eq:prim}
\end{equation}
For all cases in Fig.~\ref{fig:signal_all}, $\nu$ is set at 10 GHz. 
\subsection{Summary of other CP sources}
\label{sec:summary}
Under the realms of the Standard model, the CMB does not have any
significant intrinsic Stokes V component. However, an early universe
symmetry-breaking, or the presence of new-physics, or scattering
processes in presence of the primordial magnetic field could induce an
intrinsic Stokes V component in the CMB. Many of these possibilities are
summarized in Table.~(\ref{table:table1}) along with references. 
One of the most concerning source of non-cosmological 
CP is synchrotron emission in our galaxy. GSE is intrinsically
circularly polarized and therefore poses as a foreground to cosmologically important CP sources in
the CMB. The GSE is produced by the cosmic ray
electrons in presence of the galactic magnetic field. We will discuss
the GSE in much more detail in
Sec.~\ref{sec:foregroundsa} and Sec.~\ref{sec:foregroundsb}.
Table.~(\ref{table:table1}) also lists sources for FC driven and intrinsic 
CP generation. 
In each case, Table.~(\ref{table:table1}) lists frequency and magnetic
field dependence of the signal. Each case also projects an expected
signal at $\nu$=10 GHz.
\section{Modeling of foregrounds to the CMB CP}
\label{sec:foregrounds}
\subsection{Intrinsic CP of the GSE}
\label{sec:foregroundsa}
Total brightness of the galactic radio sky is dominated by the diffused
synchrotron emission due to the 
cosmic ray electrons and positrons gyrating in the
GMF \citep{cowsik}. 
This emission is significant between
frequencies of a few tens of MHz to a few tens of GHz. The primary sources 
of cosmic ray electrons are supernovae remnants, pulsars in the galaxy.
The energy range of the cosmic ray electrons are between a few hundred
Mev to tens of Gev. The GMF  strength $\sim$ a few
$\mu$G. The ordered components of the GMF is
coherent on kpc scales, however, the GMF also has a random component 
which varies in scales of a few hundred parsecs \citep{beckreview}.

Synchrotron emission
due to the MW galaxy happens to be the strongest source of foreground to
the CMB observations at low frequencies \citep{wmapy1fore,wmapy7fore}. Synchrotron emission is also
naturally expected to have a rather high level of LP, and there has not be
any significant measurements of its CP. Synchrotron
emission coming from relativistic cosmic rays, 
is expected to be elliptically polarized even with an 
isotropic distribution of the electron velocity \citep{westfold}.
Emissivity, $\eta_V$ or the 
intrinsic CP of the GSE  under a power-law
electron energy distribution is given by 
\begin{equation}
dV^{\rm sync} = F_1\cot\theta\left(\frac{\nu}{\nu_{B_{\perp}}}\right)^{-1/2}
\eta^{\rm sync}_{I}dz
\label{eq:vsync}
\end{equation}
where $dz$ is the elemental length along the line of sight. $\eta_I$ 
is the emissivity associated to the unpolarized intensity of the synchrotron
emission. $\nu_{B_{\perp}}$ is the gyro-frequency of the magnetic field
component perpendicular to the line of sight. $F_1$ is a function of the spectral index of the electron energy
distribution and its detailed form is found in~\cite{westfold}. 
In this paper, for simplicity, we set $F_1 =1.37$ (see eq.~(31-35) of \cite{westfold})
which is obtained using reasonable 
parameters for the electron energy distribution obtained 
from the observation of the Crab nebula \citep{crab,crab2} and low
frequency observations from the Planck satellite \citep{planckdiffuse}. Typically,
F$_1$ lies between 1 and 2, if 1<$\alpha_{\rm sync}$<3, $\alpha_{\rm
sync}$ being the spectral index of cosmic ray electron
energy distribution. 

\subsection{CP generation via FC in the MW galaxy}
\label{sec:foregroundsb}
GSE is a crucial and
significant source of foreground towards cosmologically important
sources of CP in the CMB. 
In this paper, we only focus on the intrinsic emission
of the galactic synchrotron as the most significant foreground
towards the cosmic CP. Circularly polarized
synchrotron emission in the MW 
galaxy is described by the term $\eta_V$
in Eq.~(\ref{eq:transfer}) and its precise form is given in
Eq.~(\ref{eq:vsync}).

Another possible mechanism of galactic CP generation is the FC in the galaxy due to
relativistic cosmic ray electrons gyrating in
the GMF. 
In this section, using simple analytic models, 
we will estimate the magnitude of FC induced galactic CP.
We begin with the following cosmic ray electron density
distribution \citep{page,strong}. 
\begin{equation}
N_{\rm cre}d\gamma=C_{\rm cre}\exp(-r/h_r)sech^2(-h/h_z)\gamma^{-p}d\gamma
\label{eq:electron}
\end{equation}
where r  is the galactic radius and z is the height. Synchrotron
spectral index, $p$ $\sim$ 3, Lorentz factor,
$\gamma$ lies between 100 and 300 for the galaxy and 
C$_{\rm cre}$=4 $\times$ $10^{-3}$ cm$^{-3}$. The
radial scale and disc height are respectively 
set by $h_r=5$ kpc and $h_z$=1 kpc. We use $r \sim $ 0.5 kpc 
and $h_z \sim$ 0.5 kpc at $\nu$=10 GHz in Eq.~(\ref{eq:electron}) to obtain relativistic electron density
in the MW galaxy. We then use Appendix D. of \cite{beckert} to obtain 
FC coefficient in the galaxy, $\kappa^{\rm galaxy}_C \sim 10^{-15}$kpc$^{-1}$.

The amount of CP induced via the FC effect in the galaxy is $\sim$ $\kappa_CU_{\nu}$, where $U_{\nu}$
is the total incoming LP due to the CMB and the GSE at a given frequency $\nu$. $U^{\rm CMB}\sim 10^{-6}$K, and
therefore induces a CP due its passage through the galaxy as $\delta
V^{\rm CMB}_{\rm galactic~FC} \sim \kappa^{\rm galaxy}_{C}U_{\nu}^{\rm
CMB} \sim 10^{-21}\rm kpc^{-1}$K, which is much smaller
than any other cosmologically important source of CP in the CMB.

GSE also has a significant level of intrinsic LP \citep{westfold} 
and therefore 
is subjected to FC effects in the
galaxy. In Sec.~\ref{sec:psm}, we show the 
expected level of LP in the
GSE (see Fig.~\ref{fig:planck_vs_hammurabi}) at $\nu$=30 GHz. We note
that the highest level of intrinsic LP of the GSE 
is $\sim$ a few tenths of K. This level does not change significantly
with our frequencies of interest ($\nu$$>$ 5 GHz). Therefore, CP
induced in the galaxy due FC of the GSE is $\sim$ $\kappa_{C}^{\rm
galaxy}U_{\nu}^{\rm GSE}$ $\sim$ 10$^{-16}$kpc$^{-1}$K. 
In contrast, intrinsic CP of the GSE at
$\nu$=10 GHz is $\sim$ $10^{-6}\rm kpc^{-1}$ K, obtained using equations
in Appendix D. of \cite{beckert} with B $\sim$ 10$\mu$G. This level is
also supported by Fig.~\ref{fig:hammu_map}. Therefore, FC induced CP
of the GSE is negligible compared to its intrinsic CP.

CP levels in the galaxy is a function of frequency and the relative
importance of different channels to generate CP, depends on the
frequency. The $frequency$ $dependence$ of the ratio of intrinsically generated CP vs FC induced CP in
the galaxy is given by
\begin{equation}
\frac{\eta_V}{\kappa_C U_{\nu}} \propto \nu^2 \cot \theta
\log \left(\frac{\nu}{\nu_B\gamma_{\rm min}^2}\right)^{-1} U_{\nu, \rm
GSE}^{-1}.
\label{eq:vvsu}
\end{equation}
where $U_{\nu}$ is the total incoming LP in the galaxy, composed of
contributions from both the CMB and the GSE. $\theta$ is the angle
between the magnetic field and the line of sight, $\gamma_{\rm min}$ is
the minimum Lorentz factor for the relativistic electrons in the galaxy
and $\nu_{\rm B}$ is the cyclotron frequency given by $\nu_{\rm B}=2.8
(B/1\mu G)$. The ratio $\eta_V/\kappa_CU_{\nu}$ is $\sim
10^{9}/U_{\nu}(K)$ at
$\nu$=10 GHz, B=10$\mu$G, $\gamma_{\rm min}=100$ and $\theta=\pi/4$. LP
in the CMB is not a function of frequency and is given by
$U_{\nu}^{\rm CMB} \sim 10^{-6}$ K. The level of LP in the galactic
synchrotron emission is significant compared to the unpolarized
intensity. 
$U_{\nu}^{\rm sync} <0.2$ K at $\nu$=10 GHz and
eventually falls off with higher frequency. Therefore, 
along a given line of sight, $\eta_V/\kappa_CU_{\nu}$ is a monotonically
increasing function of frequency. Therefore, in our frequencies of
interest (5 GHz <$\nu$ <30 GHz) for the CMB CP measurement, FC induced
CP of the CMB or the GSE is not an important effect.

To every
emission there is an associated absorption. This applies to both the
unpolarized Stokes intensity, I and circular polarization Stokes
 intensity V. The intrinsic emission in Stokes V will be extinct if the
emission and absorption were perfectly balanced, or  
$(\eta_V-\kappa_V I) \sim 0$. Therefore, it is also important to consider the absorption of the circularly
polarized emission in the galaxy.  
The absorption is given by $\kappa_V I_{\nu}$.
Following Appendix D of \cite{beckert} we obtain,
\begin{equation}
\frac{\eta_V}{\kappa_V I_{\nu}}=m_e \nu^2
 \left(\frac{\nu}{\nu_{B_{\perp}}}\right)^{1/2} \psi(p) /I_{\nu}
\label{eq:vvsi}
\end{equation}
where $\psi(p)$ is a function of the spectral index, $p$, of
the relativistic electron energy distribution.
For the galaxy, we use $p \rightarrow 3$. 
For the GSE, $I_{\nu} \sim \nu^{-(p-1)/2}$.
The ratio $\eta_V/(\kappa_V I_{\nu})$ is $\sim 10^{14}$ at $\nu$=10 GHz,
B $\sim$ 10 $\mu$G, $I_{\nu} \sim 10^{-2}$ K and is an increasing
function of frequency. This implies that absorption of the CP emission
in CMB observation frequencies (1 GHz or above) is not significant.
Absorption of the synchrotron emission component is however significant 
in so-called self-absorbed
synchrotron sources where $(\eta_V-\kappa_VI )\rightarrow 0$. This
scenario is realized at much
lower frequencies, $\nu_{\rm self}$ $\le$ 10 MHz \citep{smootsync}.
Therefore, synchrotron self absorption of its circularly polarized
emission is not a concern in our case.

Synchrotron self-absorption of the unpolarized intensity, 
is also not important for the galaxy in the
frequencies of interest ($\nu$ $>$ 1 GHz). In the case of very low frequencies $<$ 1 GHz, some
extra-galactic sources could become self-absorbed or optically thick.
In this low frequency
regime, synchrotron flux from the sources decreases with decreasing
frequency. On the contrary, at higher frequencies, flux emitted by the synchrotron
sources decreases with increasing frequency. This turn-over in
flux-frequency relation pollutes 
the smooth synchrotron frequency dependence, altering the spectral index
of the synchrotron brightness temperature. Spectral smoothness is
important in order to successfully remove foregrounds. 
Synchrotron flux from the
galaxy is still high at frequencies $\ge$ 1 GHz. However, due to the
smooth dependence of the synchrotron flux on the frequency, foreground
removal via a polynomial fit, is easier. 
This is especially relevant where the signal of
interest (for example, the CMB CP due to galaxy clusters) 
is lower than the foregrounds. Unless
the number of such sources is small enough, it is wiser to confine the 
search for
the cosmic CP at frequencies $\ge$ 5 GHz. This is also the motivation
for us to confine the CMB CP observation frequencies between 5-30 GHz.

Below we summarize the conclusions from the current section.
\begin{itemize}[label={--}]
\item CP induced in the CMB due to the MW galaxy (via the FC
mechanism) is much smaller than
the levels of CP induced in the CMB due to cosmologically important
sources (see Table.~(\ref{table:table1})).
\item CP induced in the GSE via the FC
mechanism is much smaller than the intrinsic emission of circularly
polarized synchrotron radiation in the galaxy.
\item CP induced intrinsically (via the $\eta_V$ term in
Eq.~(\ref{eq:transfer})) is higher than the FC
induced CP in the galaxy at all frequencies of interest.
\item Self-absorption of the circularly polarized GSE is not important in the frequencies of
interest.
\item Self-absorption of unpolarized intensity of the 
GSE  may pollute the smoothness of 
synchrotron spectra at
frequencies $<$ 1 GHz. 
\end{itemize}

\begin{figure}
\begin{tabular}{cc}
\includegraphics[height=0.3\textwidth]{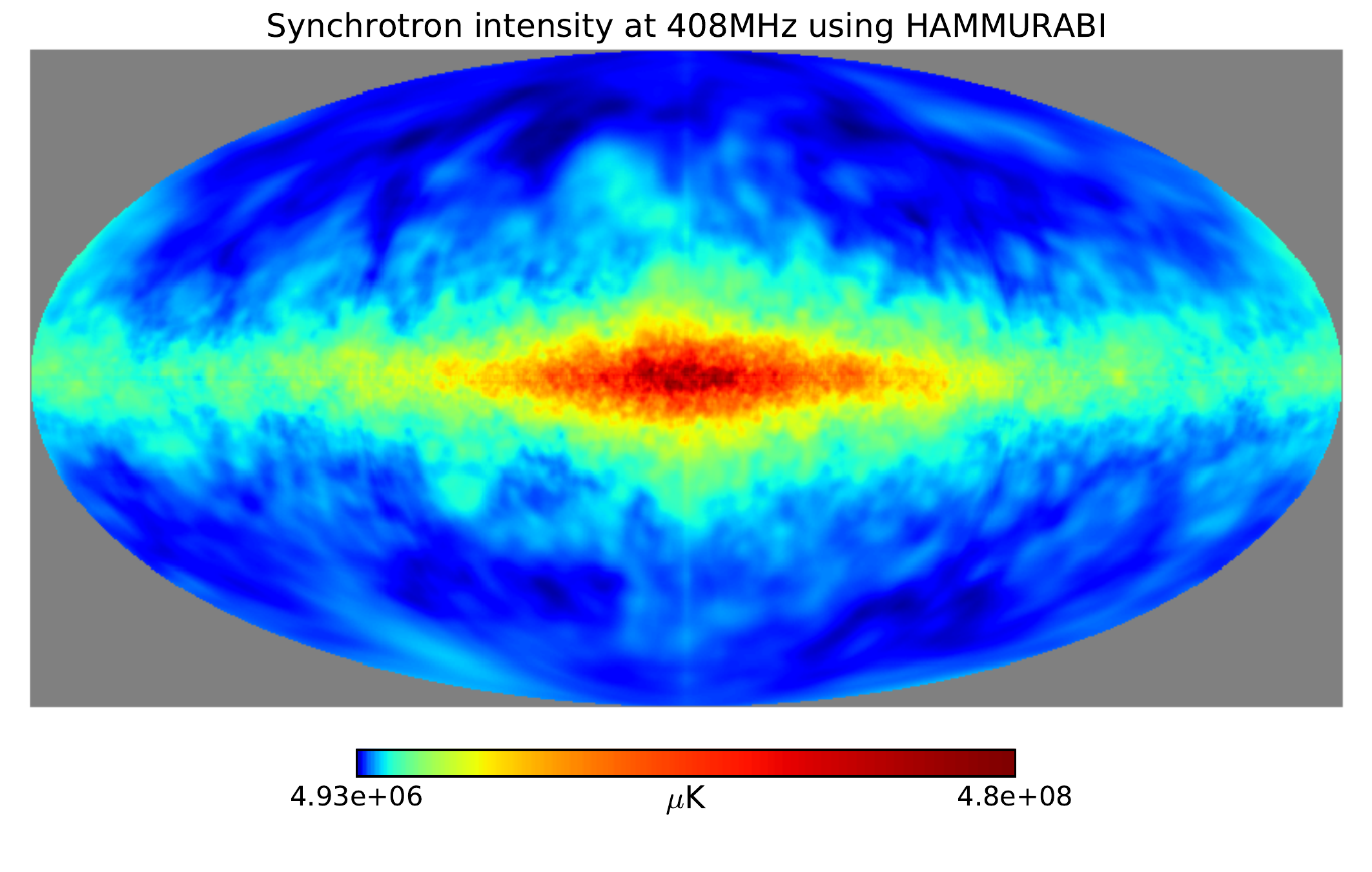}
&\includegraphics[height=0.3\textwidth]{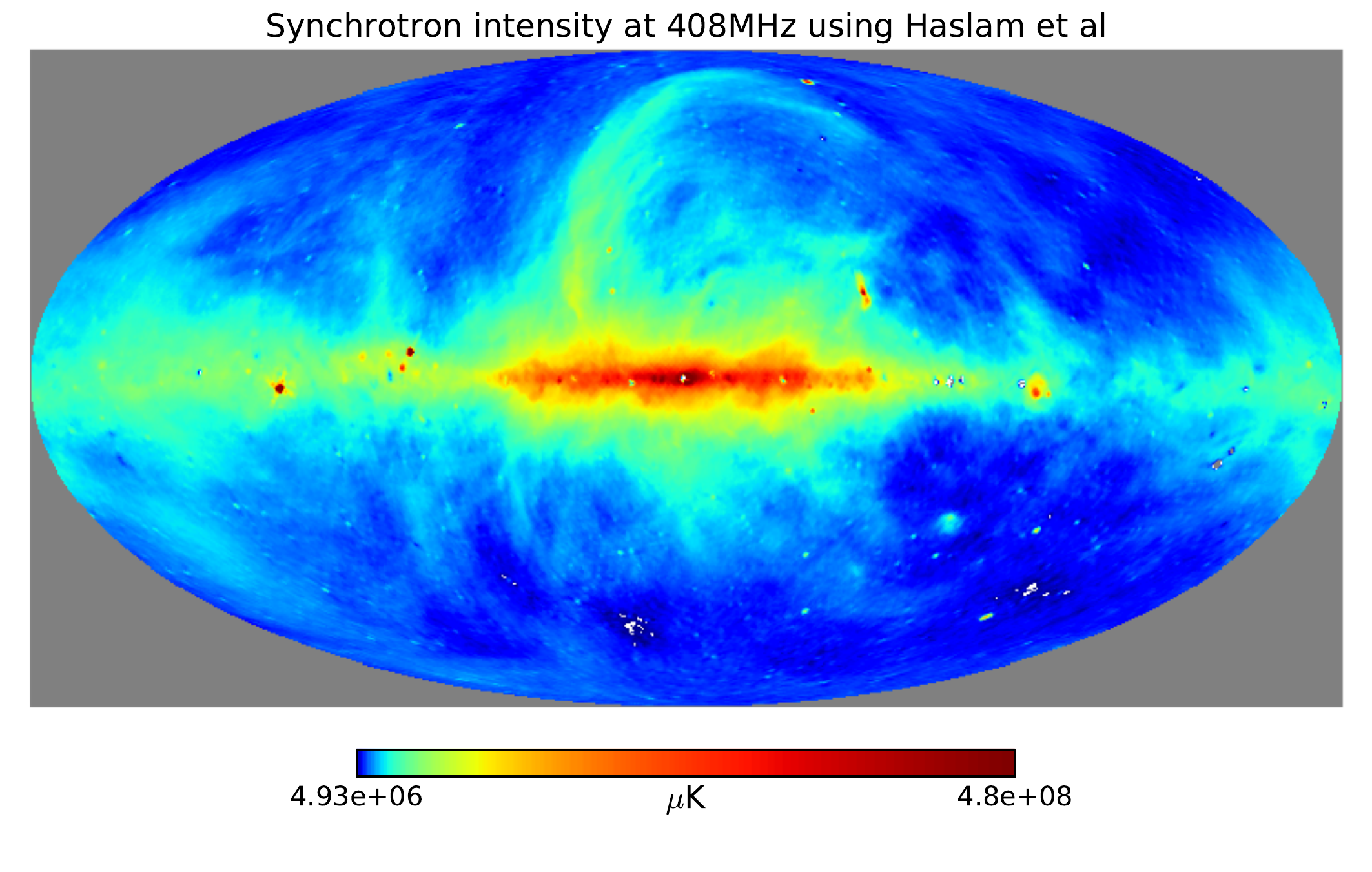} \\
\end{tabular}
\caption{Comparison between $\hammu$ \citep{waelkens} generated map of galactic
synchrotron intensity at 408 MHz and \cite{rema} generated 408 MHz
map derived from the \cite{haslam} data. Intensity in both maps are
presented in Rayleigh-Jeans temperature units equivalent to $T_{\rm
sync}$=I$_{\nu}$c$^2$/(2k$_{\rm B}\nu^2$). The maps were smoothed at a
resolution of 1$\degree$. Maps with a grey background in this paper
represent maps in logarithmic temperature scale. 
}
\label{fig:haslam_vs_hammurabi}
\end{figure}

\begin{figure}
\begin{tabular}{cc}
\includegraphics[height=0.35\textwidth]{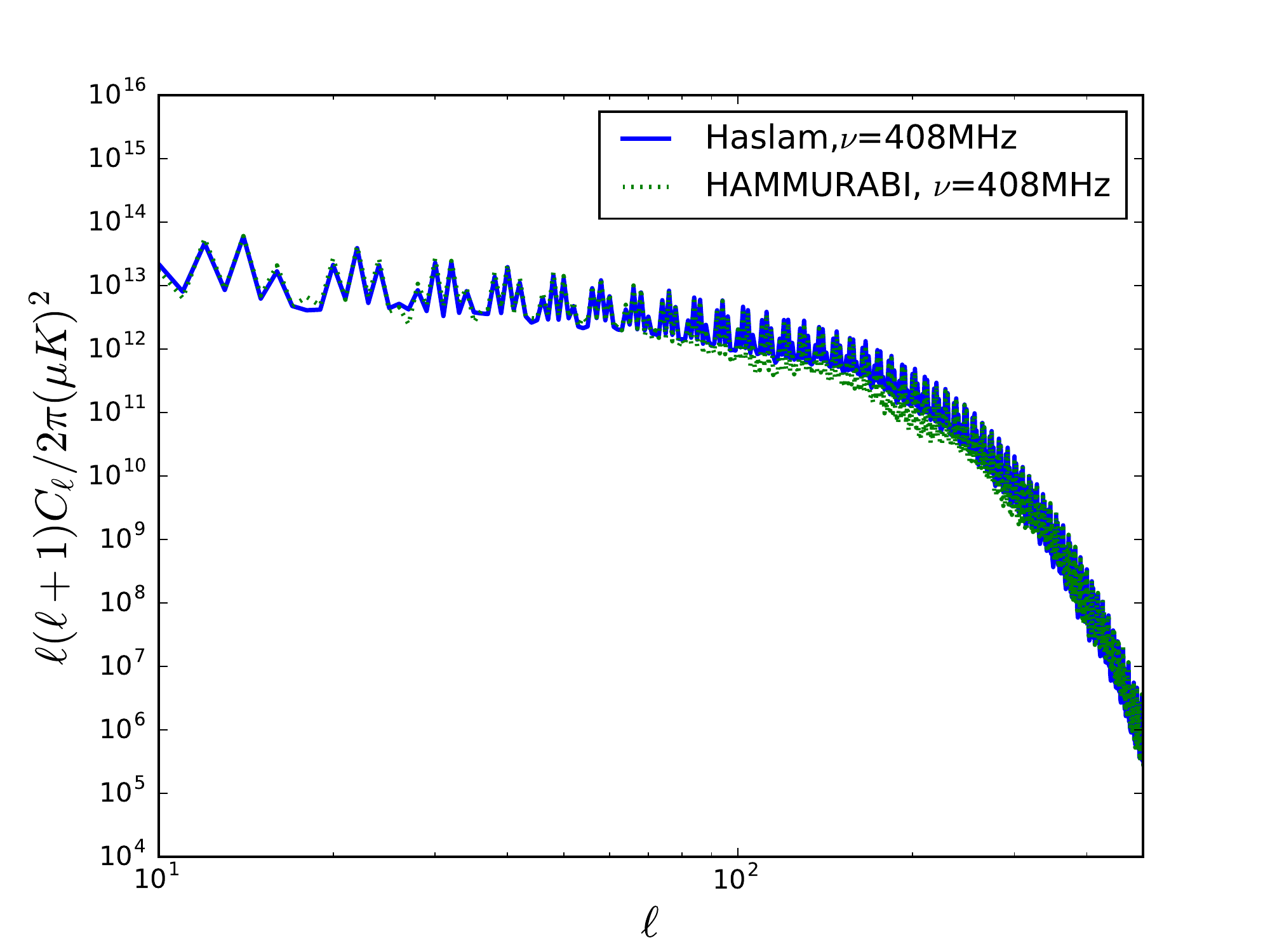}
&\includegraphics[height=0.35\textwidth]{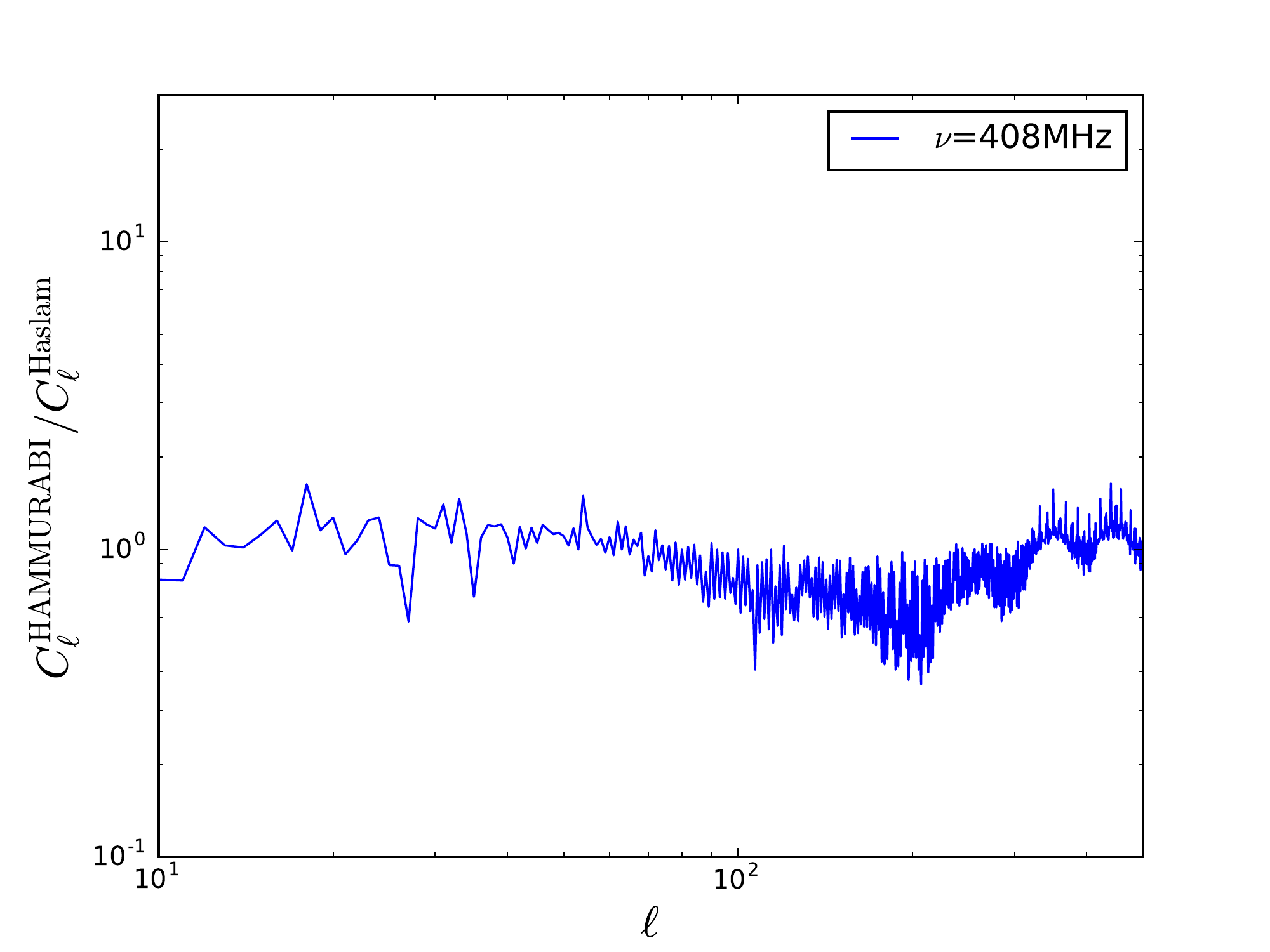} \\
\end{tabular}
\caption{Comparison between the angular power spectra of $\hammu$
\citep{waelkens} generated map of galactic
synchrotron intensity (Stokes I) at 408 MHz and \cite{rema} generated 408 MHz
map using the Haslam \cite{haslam} data. The power spectra were scaled to match
at $\ell$=100 which corresponds to approximately 1$\degree$ resolution.
Intensity maps were computed in Rayleigh-Jeans temperature equivalent
units.}
\label{fig:cl_haslam_vs_hammurabi}
\end{figure}

\section{Non-synchrotron foreground sources of circular polarization}
\label{sec:foregroundsc}
The MW galaxy is bright in free-free emission in our frequencies of
interest, 5-30 GHz. The free-free signal does not attribute to polarized
sky. There is some spinning dust signal which is 
linear polarized \citep{planckdiffuse}, however,
linear polarization in these frequencies due to spinning dust is much smaller
than LP due to synchrotron. An intrinsic CP due to spinning dust is not expected. There is also little
chance of FC of the LP (due to dust), as we have seen
in Sec.~\ref{sec:foregroundsb} that the FC coefficient
 due to the MW galaxy is
very small. Therefore, we do not expect any appreciable CP due to
the free-free and the spinning dust emission in the galaxy, in the frequencies
between 5-30 GHz.   

\cite{oxygen} discusses recent observations of meso-spheric oxygen induced
CP at large angular scales. Temperature equivalent
CP due to the atmospheric source is sensitive to
the height of the atmospheric sources. Typically, this signal is
described by a dipole-like large scale structure with 
$\rm V \sim (15-100) \times 10^{-3} \mu$K.
We will neglect this atmospheric 
contribution into the noise estimate. Oxygen 
related effects are only important in the largest 
scales for the balloon based experiments. They are however, more
serious for the ground based experiments that are 
aiming to detect a circularly polarized component in the CMB.

Generally, brightness in free-free emission or dust emission will not
affect the CP measurement unless there is leakage, which causes mixing
between different types of signal, such as polarized signal and
non-polarized signal. In this paper, we neglect leakage of any kind.
 
\section{Construction of the Stokes V maps due to the GSE}
\label{sec:maps}
\subsection{The $\hammu$ code: Implementation of CP due to the GSE}
\label{sec:hammu}
We use the $\hammu$ code
(\cite{waelkens}) to create maps of Stokes parameters I, Q, U and V
due to the GSE. Calculation 
of Stokes I, Q and U
parameters are part of the original implementation of $\hammu$ and 
clearly described in \cite{waelkens}.
These simulations use an input magnetic
field, free electron density and relativistic electron density
models to output the Stokes vectors into Healpix formatted maps
\citep{healpix} at a given frequency and spatial resolution. Below we
summarize the precise inputs used for the $\hammu$ code to generate the
Stokes V simulated maps.

\begin{figure}
\begin{tabular}{cc}
\includegraphics[height=0.3\textwidth]{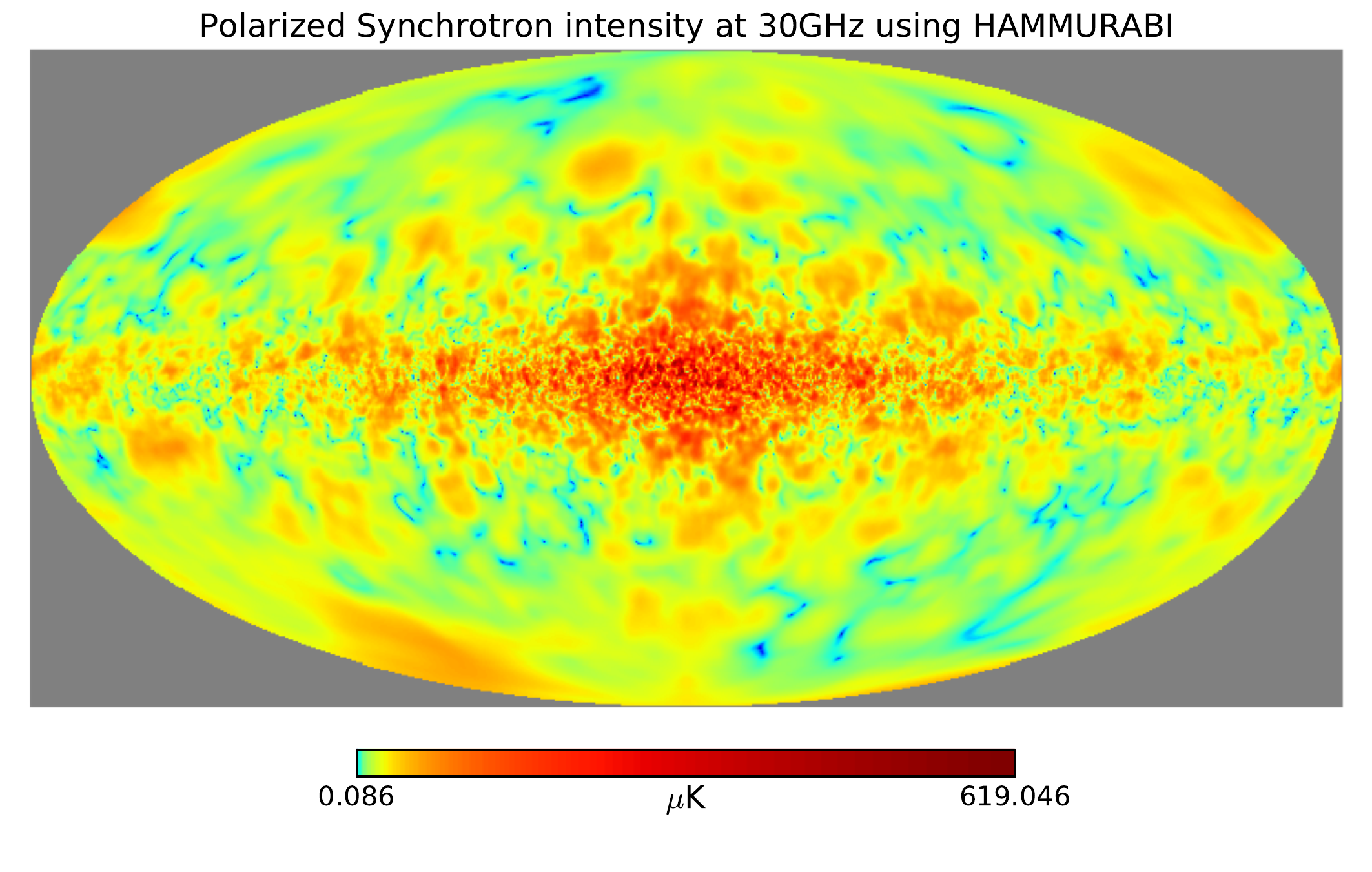}
&\includegraphics[height=0.3\textwidth]{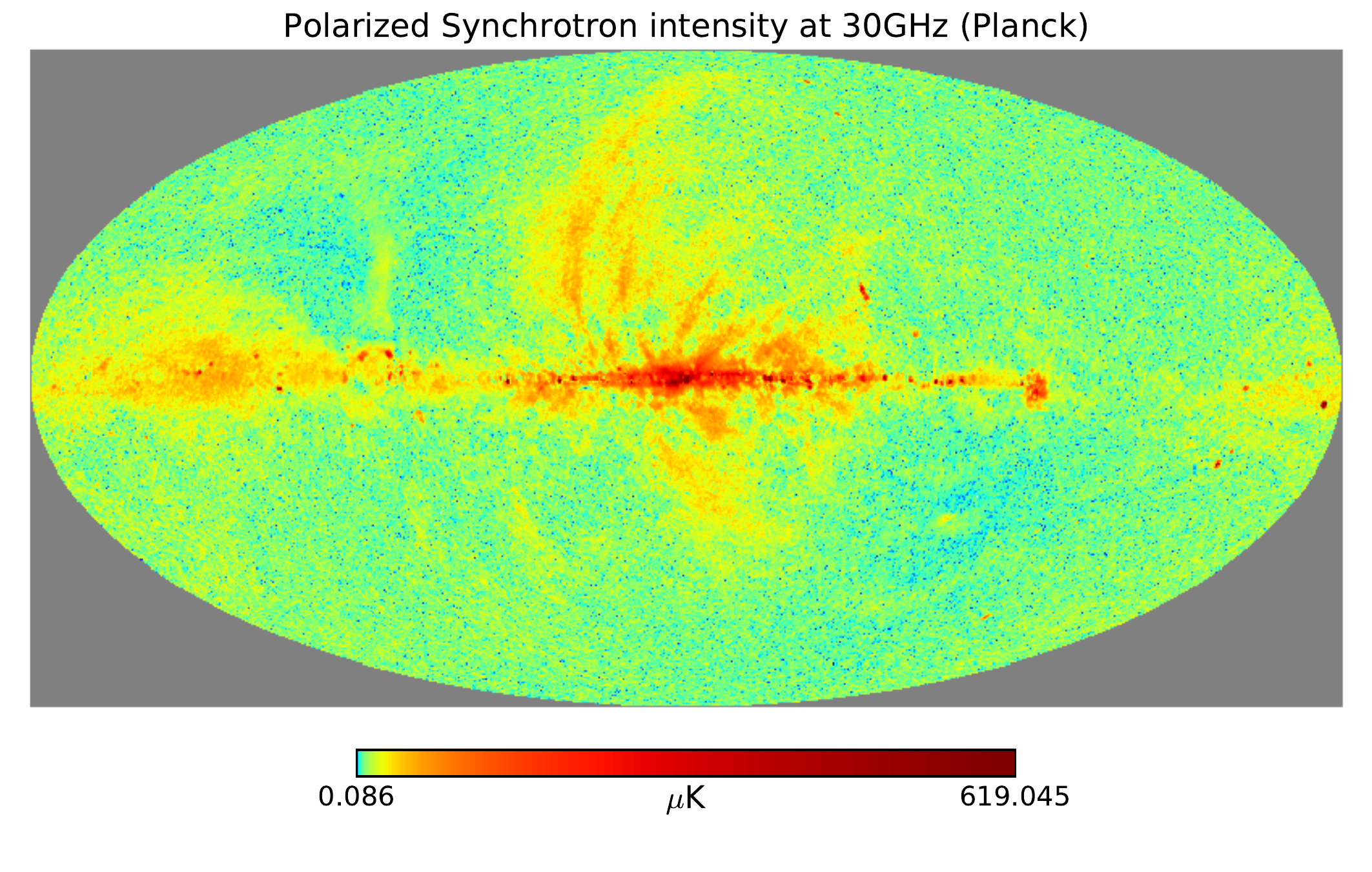} \\
\end{tabular}
\caption{Comparison between $\hammu$ \citep{waelkens} generated map of polarized galactic
synchrotron intensity at 30 GHz and 30 GHz
map of the polarized galactic synchrotron intensity derived from
the \cite{planckgeneral}
data. The maps were smoothed at 40 arcmin resolution. The polarized
intensity is derived as P=$\sqrt{Q^2+U^2}$. Maps are expressed in
Rayleigh-Jeans temperature equivalent units and the maps with a grey
background represent logarithmic temperatures scale.
}
\label{fig:planck_vs_hammurabi}
\end{figure}

\begin{itemize}[label={--}]
\item A 3D GMF model composed of both large scale
field and turbulent component. Large scale field is coherent over
scales $\sim$ kpc and is described in \cite{sun}. The turbulent
component is described in \cite{han} and is coherent over scales of few
hundred parsecs.
\item A 3D model of cosmic ray electron density given in \cite{sun}.
\item A 3D thermal electron density model described in \cite{ne2001}.
\end{itemize}

Please note, Stokes V calculation is not part of the original $\hammu$
implementation. We use Eq.~(\ref{eq:vsync}) to implement the construction of Stokes V field
due to intrinsic synchrotron emission of the MW galaxy,
into $\hammu$. All the Stokes vector outputs are expressed in
temperature units of K and Healpix formatted maps of user specified
resolution. A synchrotron spectral index of p=2.8 was used in all
$\hammu$ simulation unless specified otherwise. 

\subsection{Comparison between $\hammu$ simulations and observations}
\label{sec:psm}
The main goal of this section is to justify the use of 
the $\hammu$ code in galactic CP power spectrum calculation.
In order to do so, we use the component separated synchrotron maps of
both intensity and polarization provided at the Planck collaboration
website.

The data-driven model of GSE in the Planck website is based on a few
datasets. They are,
\begin{itemize}[label={--}]
\item 408 MHz synchrotron emission map \citep{haslam}.
\item WMAP low frequency observations \citep{wmap} (with
resolution of $\sim$ 1$\degree$).
\item PLANCK low frequency observations \citep{planckdiffuse}.
\end{itemize}

Following the methods described in \cite{rema} and using the above
datasets, a 408 MHz map of the GSE is generated (see
\url{http://irsa.ipac.caltech.edu/data/Planck/release_2/all-sky-maps/foregrounds.html}). This map has a a resolution of $\sim$ 1
$\degree$. Following similar methods and datasets a 30 GHz map of
polarized synchrotron emission \citep{planckdiffuse, psmpaper} was
generated with a resolution of $\sim$ 40 arcmin.

Next question to ask is, which quantities one must compare between the $\hammu$
simulations and the observed datasets to validate the $\hammu$ code generated
Stokes V map? To answer this question, we consult Eq.~(\ref{eq:vsync}),
which describes our implementation of the intrinsic
CP of the GSE. Eq.~(\ref{eq:vsync}) indicates
that CP depends on the
I$_{\rm sync}$ at a given point in the sky, and ratio of
line of sight component and the perpendicular
component of the GMF. Generally, synchrotron emission (in polarized and
total intensity) are proportional to B$_{\perp}$ component of the GMF.
On the other hand, galactic rotation measure (RM) 
is proportional to $B_{\parallel}$ (defined along the line of sight). 
Magnitude of CP
along a line of sight is proportional to I$_{\rm sync}$ along that line
of sight but also sensitive to B$_{\perp}$/B$_{\parallel}$, which is more
uncertain to determine. 

We make comparisons between $\hammu$ and Planck, WMAP and Haslam joint
dataset. We do this both in real space and also in terms of the angular
power spectra. Lets define an all-sky intensity field I=I$_{\rm sync}$($\Omega$) and use Healpy, a python implementation of the original Healpix, to find
\begin{equation}
\tilde{a}_{LM}=\int d\Omega W(\Omega)I_{\rm sync}(\Omega)Y_{LM}^*(\Omega)
\label{eq:almtilde}
\end{equation}
where W($\Omega$) is the mask which is 0 if a pixel is masked and 1 if
it is not. We then evaluate 
\begin{equation}
\clii=\frac{1}{f_{\rm
sky}(2\ell+1)}\sum_{m=-\ell}^{m=\ell}\tilde{a}_{\ell m}^*\tilde{a}_{\ell
m}
\label{eq:clii}
\end{equation}
where f$_{\rm sky}$ is the sky fraction. $\clii$ represent the angular
power spectrum of the unpolarized intensity of the GSE.

In Fig.~\ref{fig:haslam_vs_hammurabi} we present 408 MHz Haslam
\citep{haslam, rema} dataset derived galactic synchrotron intensity, on the right
column. On the left column, we present the 408 MHz map of the GSE
created by the $\hammu$ simulation. The spectral index chosen for this
simulation was $p$=2.8. Other inputs chosen for the simulation are
described in Sec.~\ref{sec:hammu}. Both maps are smoothed at a
resolution of 1$\degree$.

In Fig.~\ref{fig:cl_haslam_vs_hammurabi} we present the angular power
spectra of the 408 MHz Haslam data driven synchrotron power spectrum, and 
that given by the $\hammu$ simulation. We follow
Eq.~(\ref{eq:clii}) to derive the power spectra. In each case, an
identical mask was used to remove the high-foreground galactic disc from
the sky. A 20$\degree$ symmetric cut around the equator along with the
WMAP K-band mask was used. More on the specifics of other masks and
their effect on the power spectra is described in
Sec.~\ref{sec:mask}. Fig.~\ref{fig:cl_haslam_vs_hammurabi} shows
that the shapes of the power spectra are similar and the ratio of power
at each angular scale fluctuates around unity. The $\hammu$ map was
scaled with the Haslam 408 MHz map at $\ell$=100.

In Fig.~\ref{fig:planck_vs_hammurabi} we present the 30 GHz synchrotron
polarized map (left panel) derived from the Haslam, Planck and WMAP data
\citep{rema,psmpaper,wmap,planckdiffuse} and the $\hammu$
\cite{waelkens} simulations (right panel). 
The total polarization from
synchrotron emission from the galaxy is plotted. The maps are presented
in Rayleigh-Jeans temperature equivalent units. The
maps represent the total polarization given by
P=$\sqrt{Q^2+U^2}$ in the units of $\mu$K. The left panel represent the
30 GHz polarized map of the GSE obtained 
with an uniform synchrotron spectral
index of $p$=2.8. The right panel represent the data-driven synchrotron
polarization map. Each map produce similar morphology although there are
many mismatches in details. The more detailed and accurate maps could be
produced by using more accurate GMF models which are not available at
the moment.  

From Fig.~\ref{fig:haslam_vs_hammurabi}-\ref{fig:planck_vs_hammurabi},
we draw the following conclusions.

\begin{itemize}[label={--}]
\item The power spectra of synchrotron intensity between  the $\hammu$
simulation and observed data at 408 MHz match at each scale with their
fluctuation
within unity.
\item The polarization of the GSE between the $\hammu$ and observed data
at 30 GHz match in overall morphology and order of magnitude estimates
for the polarization. 
\item There are many finer details of morphological mismatch between the
polarization maps between the $\hammu$ simulation and the observed data.
This mismatch arises from inadequate GMF models and models of the cosmic
ray electron density,  which can only be
improved with more data in the future.
\item The galactic disc is the highest source of synchrotron intensity
and polarization. The disc-removed angular power spectra of the
synchrotron intensity between $\hammu$ and observed data agree reasonably
well. Therefore, S/N derived using GSE angular power spectrum estimates
is expected to be reliable (See Eq.~\ref{eq:snr}).
\end{itemize}

Currently, there are no reliable observed datasets for all-sky galactic
CP. We will describe $\hammu$ code generated simulation of galactic CP
in the next section.
\subsection{All-sky maps of galactic CP simulated by $\hammu$}
\label{sec:mapsv}
\begin{figure}
\begin{tabular}{cc}
\includegraphics[height=0.3\textwidth]{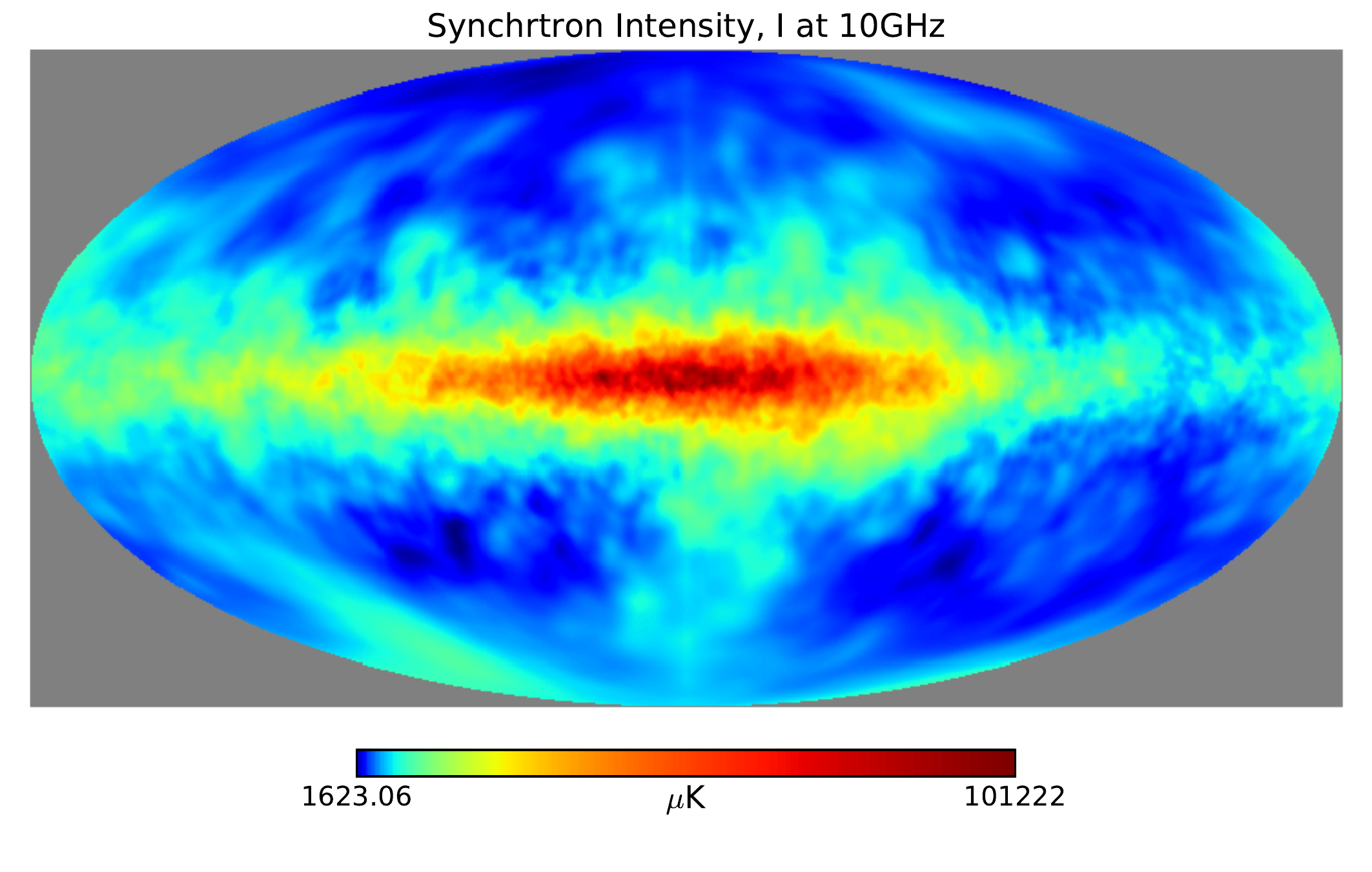}
&\includegraphics[height=0.3\textwidth]{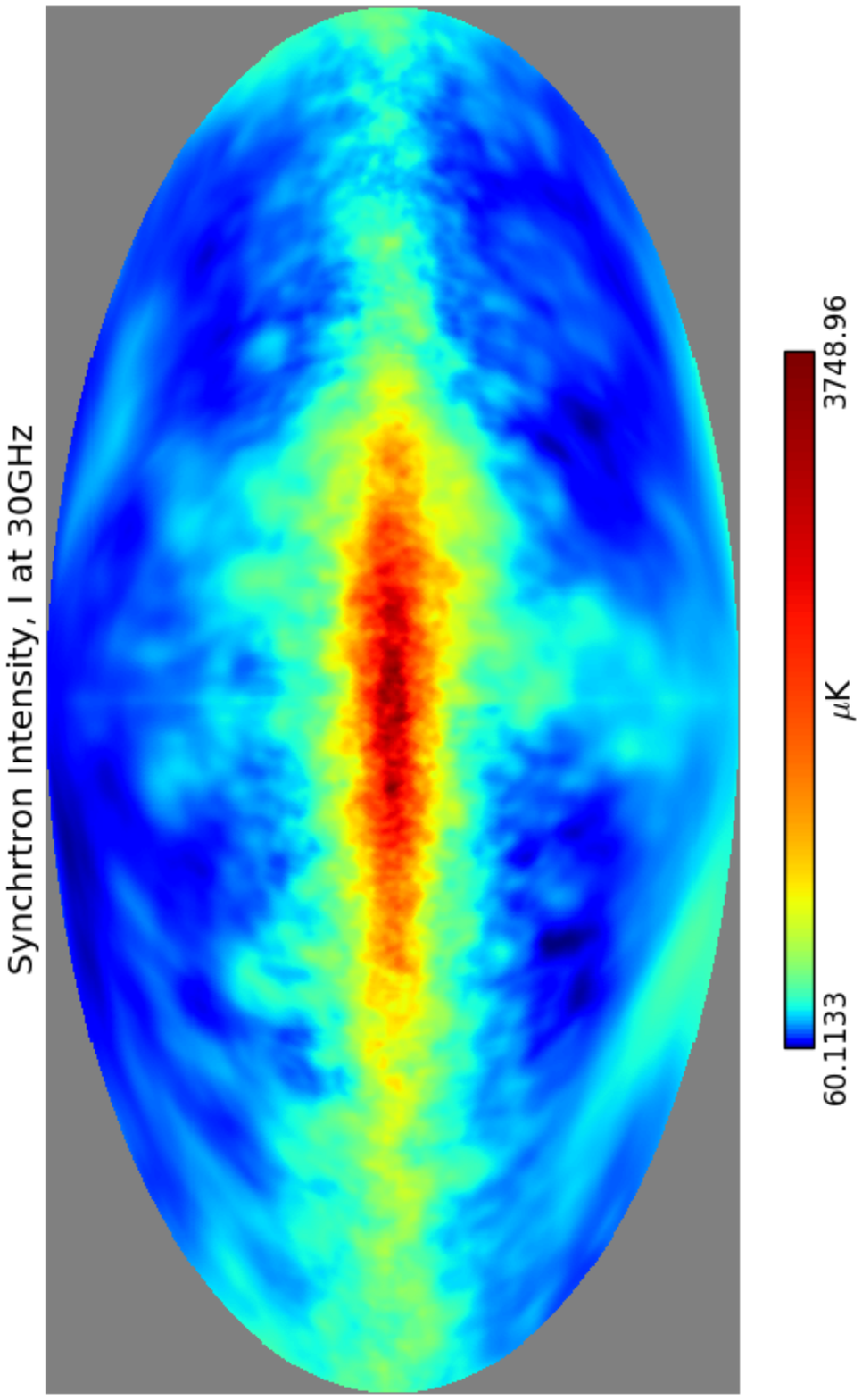}\\
\includegraphics[height=0.3\textwidth]{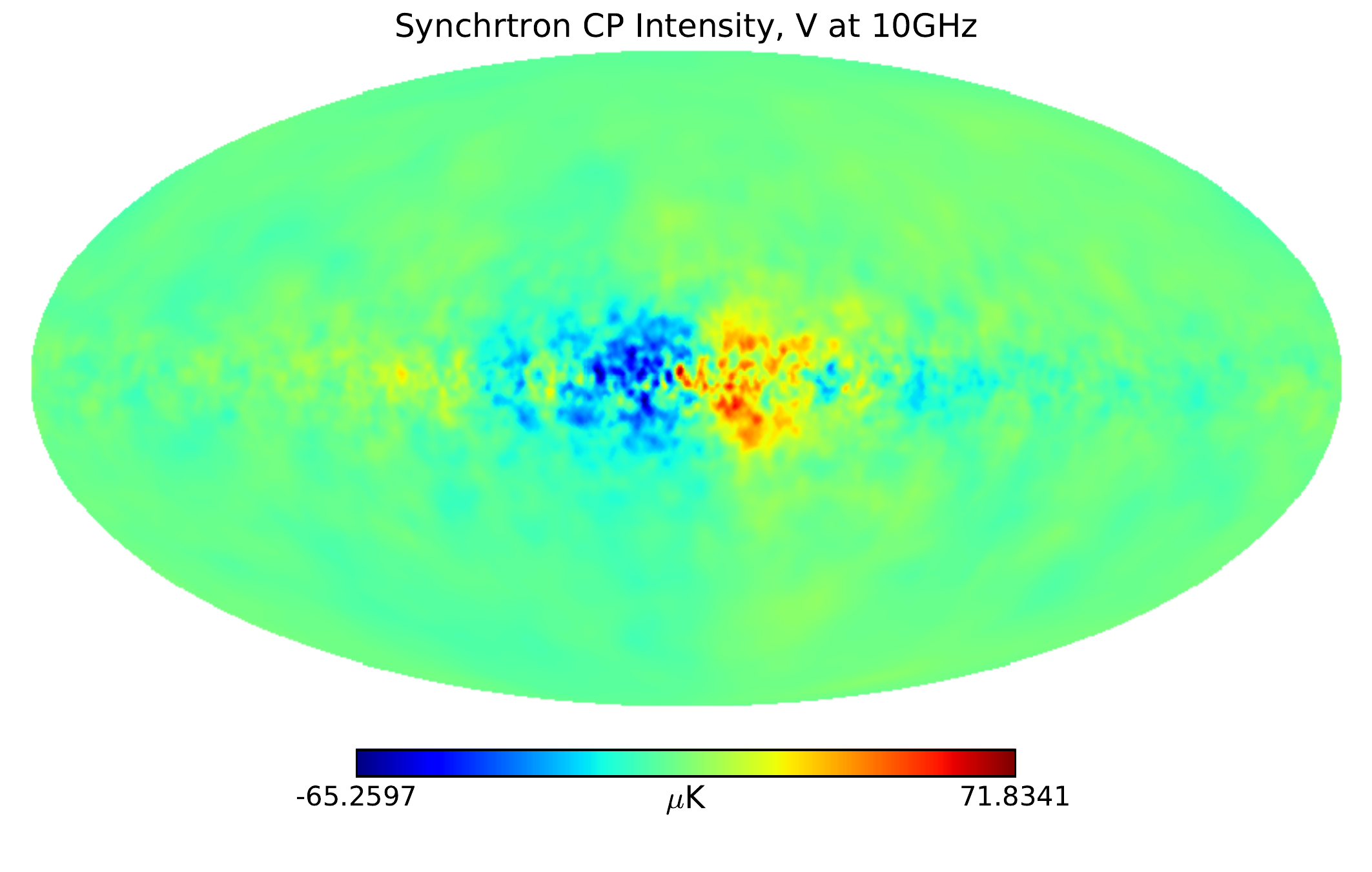}
&\includegraphics[height=0.3\textwidth]{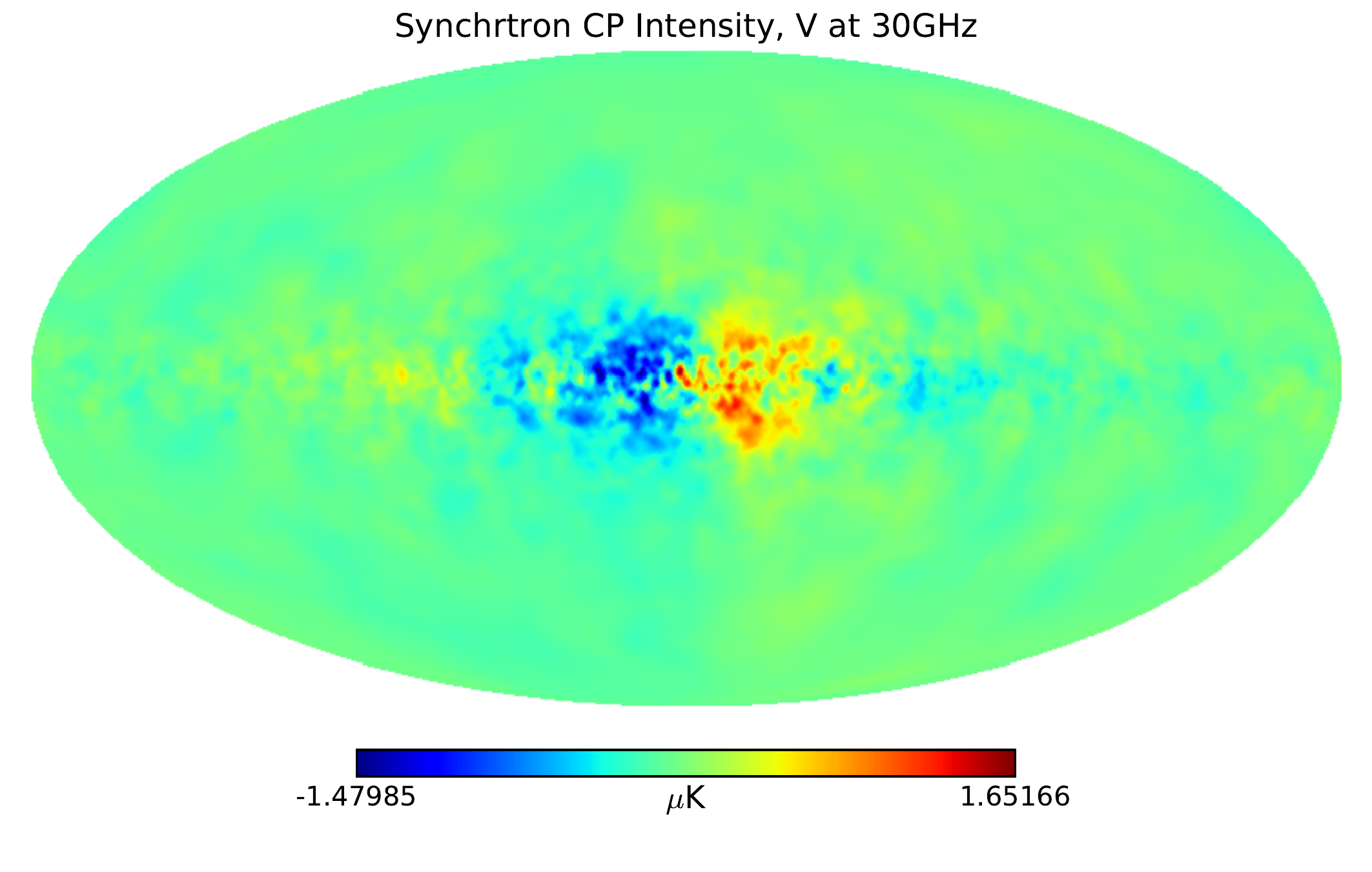}\\
\includegraphics[height=0.3\textwidth]{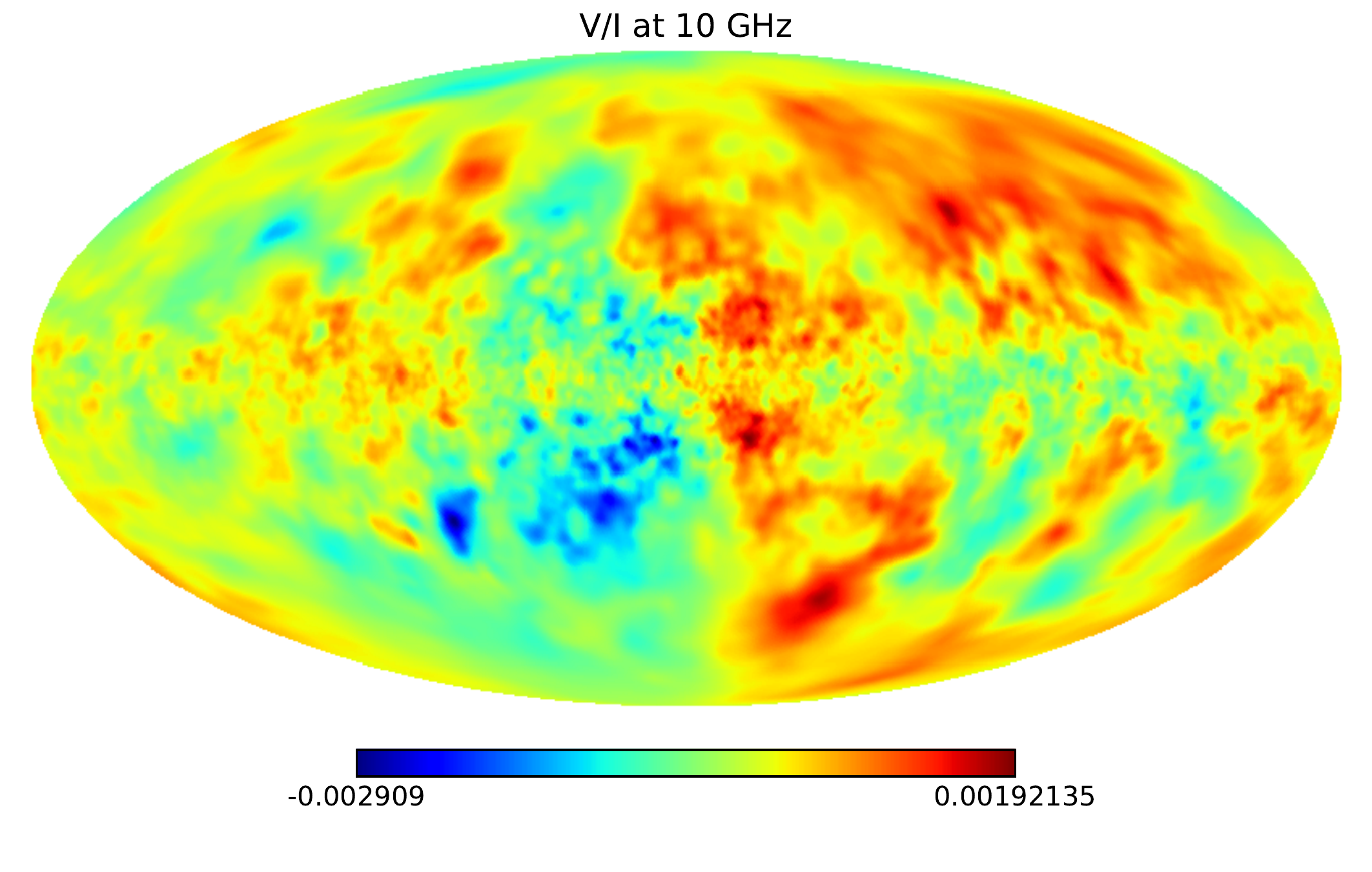}
& \includegraphics[height=0.3\textwidth]{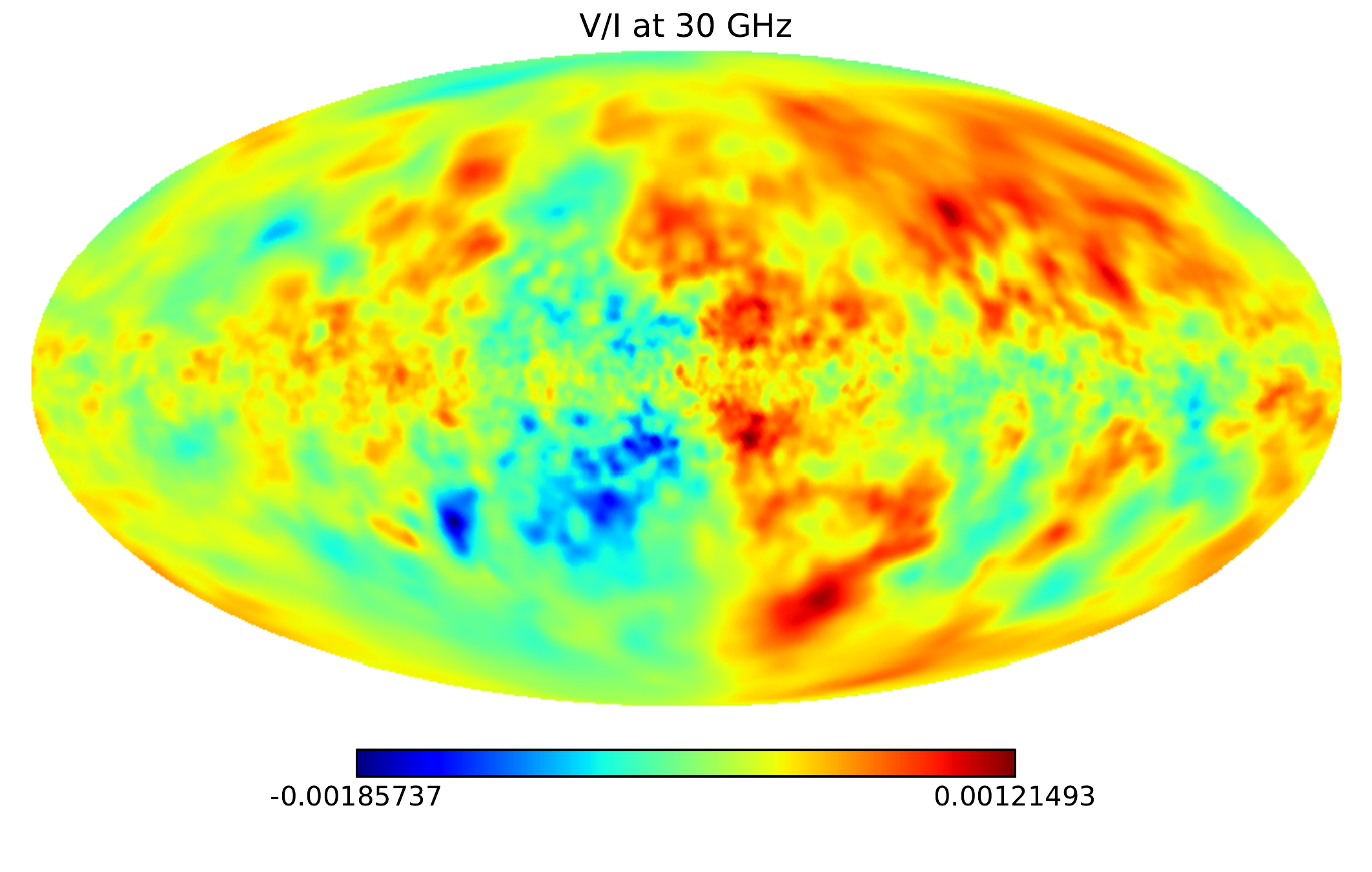}\\
\end{tabular}
\caption{Left column shows intrinsic 
Stokes I, V and V/I corresponding to the GSE at 10 GHz generated using the $\hammu$
\citep{waelkens}. Right column shows the same for 30 GHz.
The maps were generated using NSIDE=256 and then smoothed using a
Gaussian beam of FWHM $\sim$ 1$\degree$. Along a given line of sight,
each of Stokes I, V and th ratio V/I falls off as a power law with
increasing frequency. Each intensity (or polarization) map is presented
in Rayleigh-Jeans temperature equivalent units. The maps of Stokes I on the top panel are
presented using logarithmic temperature scale and have a gray background for
distinction. The order of magnitude of the Stokes I, P
(=$\sqrt{Q^2+U^2}$) and V agree well with analytic values obtained using
equations in \cite{beckert}. Stokes V intensity depends on the
frequency, $\nu$ as V$_{\rm sync}$ $\sim$ $\nu^{(-2-\alpha_{\rm sync}/2)}$, where
the spectral index, $\alpha_{\rm sync}$ $\sim$ 2.8. The maps are available upon request at
sde@ucdavis.edu or see
\href{http://somade.faculty.ucdavis.edu}{http://somade.faculty.ucdavis.edu} for
details. 
}
\label{fig:hammu_map}
\end{figure}

\begin{figure}
\begin{tabular}{cc}
\includegraphics[height=0.3\textwidth]{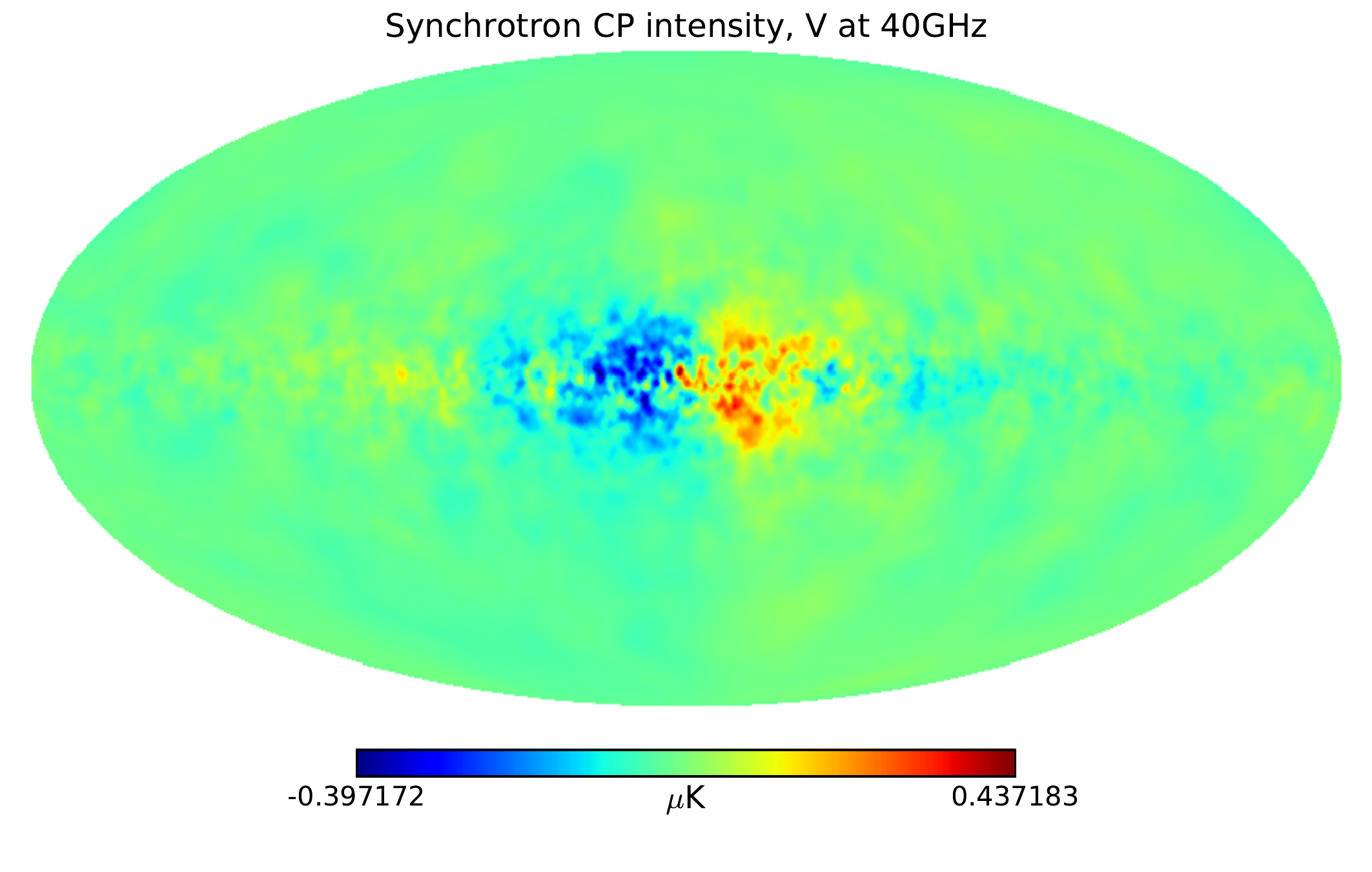}
&\includegraphics[height=0.3\textwidth]{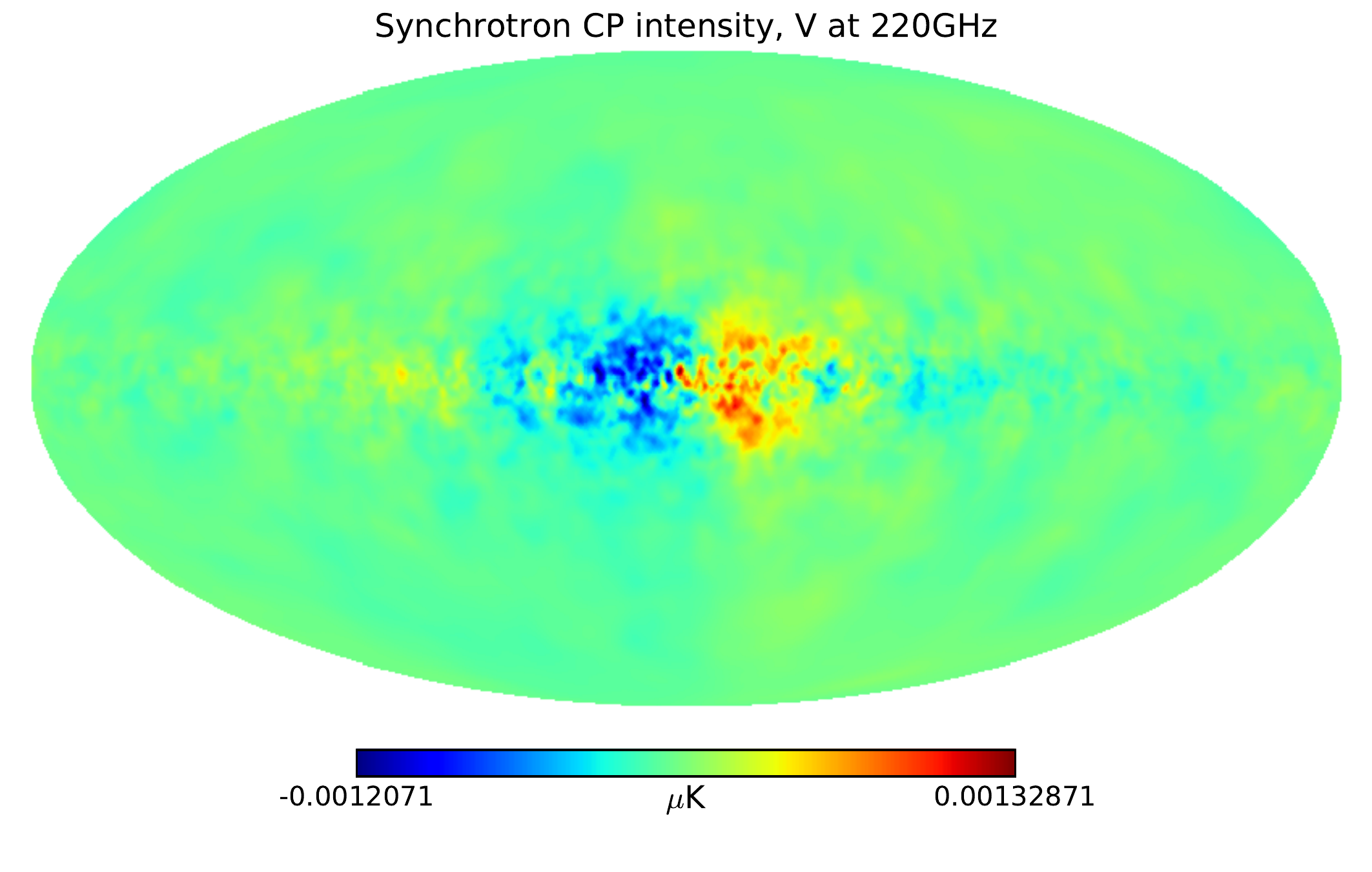}\\
\includegraphics[height=0.3\textwidth]{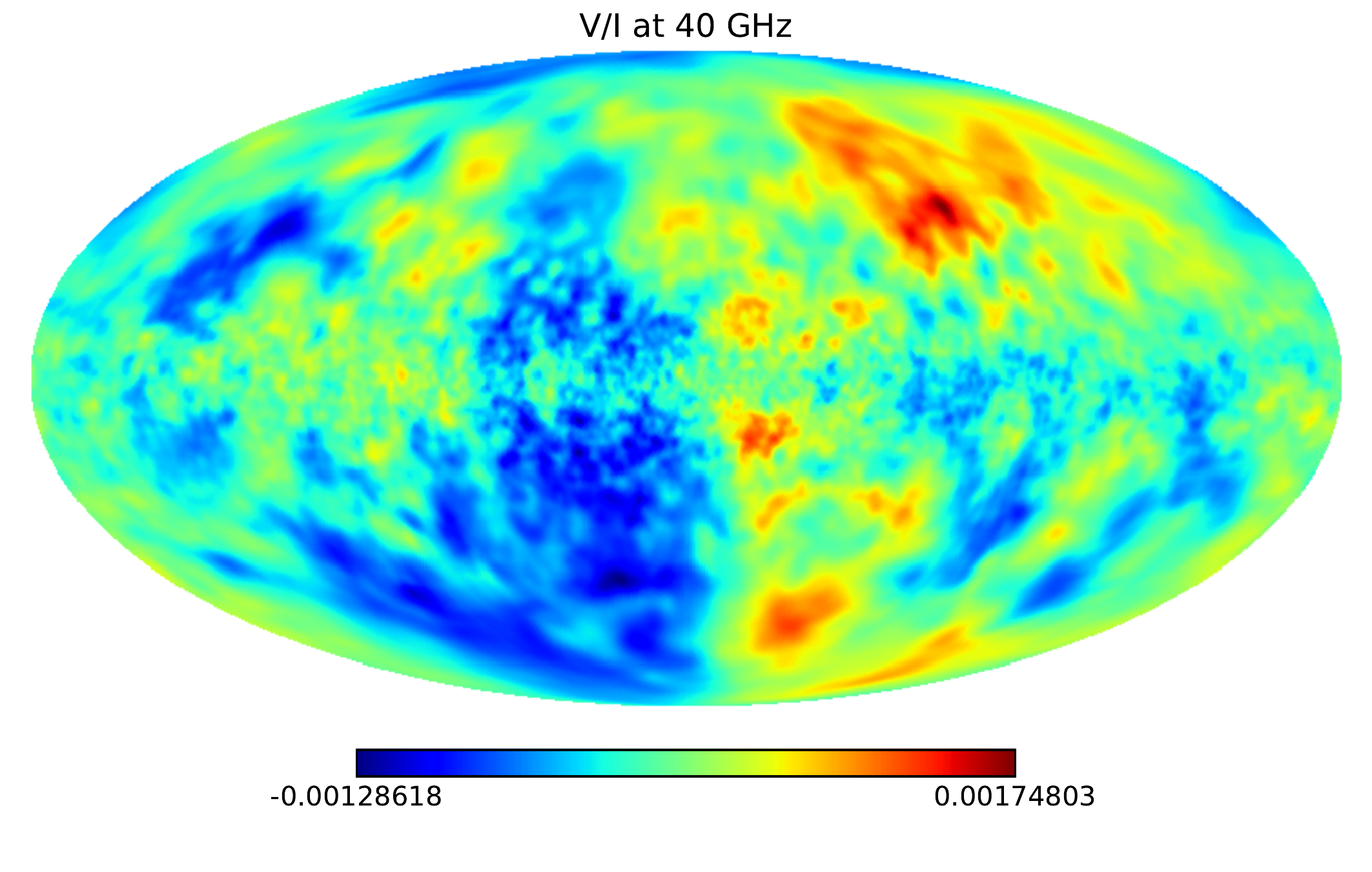}
& \includegraphics[height=0.3\textwidth]{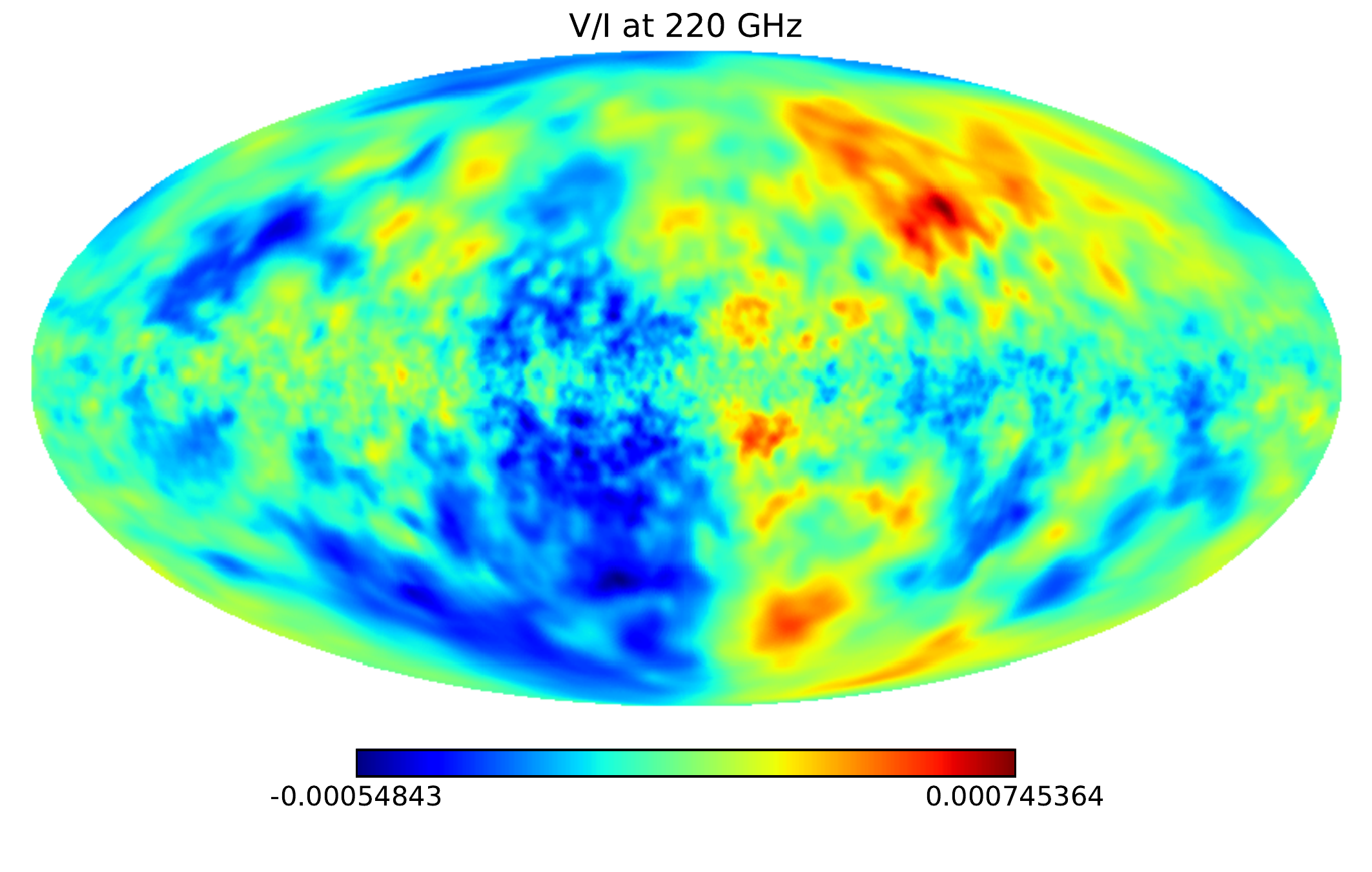}\\
\end{tabular}
\caption{
Similar to Fig.~\ref{fig:hammu_map}, except in this case the left column
represents maps at $\nu$=40 GHz and the right column represents maps at
$\nu$=220 GHz which are relevant respectively for the CLASS and PIPER
CMB telescopes. The maps are available via email at sde@ucdavis.edu or
please see
\href{http://somade.faculty.ucdavis.edu}{http://somade.faculty.ucdavis.edu}
 for details.}
\label{fig:hammu_map_class}
\end{figure}

Below, we will
describe maps of both unpolarized and polarized synchrotron emission due
to the MW galaxy, evaluated at 10 and 30 GHz. The resolution of the maps were
managed using Healpix specified parameter NSIDE, where the corresponding
angular resolution of the map is given by $\delta \theta \sim
3600'/(12~\rm NSIDE^2)$. Each map was generated with NSIDE=256.

As already discussed in Sec.~\ref{sec:foregroundsb}, FC
effects due to the MW galaxy is insignificant compared to the 
CP or the Stokes V
induced in the CMB due to
primordial effects or intrinsic V of the synchrotron emission of the MW
galaxy itself.  
Therefore, only the intrinsic generation of CP due to galactic
synchrotron emission or $\eta_V$ (in Eq.~(\ref{eq:transfer})) term was
considered for the galactic Stokes V map calculation.

In the left column of Fig.~\ref{fig:hammu_map} 
we display synchrotron radiation from the MW galaxy in Stokes I, V and
V/I. Left panel represents $\nu$=10 GHz and the right panel corresponds
to $\nu$=30 GHz. Each map was smoothed at a resolution of 1$\degree$.

In Fig.~\ref{fig:hammu_map_class} we add two sets of maps at 40 GHz and 220 GHz which
are more relevant to the upcoming CMB telescopes, CLASS and PIPER with capabilities to
measure Stokes V. Please note, that for the frequency relevant for the 
PIPER telescope, galactic CP
is lower by a factor of (40/220)$^3$=6$\times$10$^{-3}$, compared to its levels at 40 GHz which is relevant for the CLASS telescope.  

We make the following observations from the maps shown in
Fig.~\ref{fig:hammu_map} and Fig.~\ref{fig:hammu_map_class}.

\begin{itemize}[label={--}]
\item The magnetic 
field strengths are highest
around the disc of the galaxy causing the highest synchrotron signal in
Stokes I along the disc. 
\item Along a given line of sight, synchrotron intensity falls off as a
power law with increasing frequency as $\sim$ $\nu^{(-2-\alpha_{\rm
sync}/2)}$. Note that, this frequency dependence is different from
$\nu^{-3}$ dependence in the case of FC generated CP.
\item It follows from Eq.~(\ref{eq:vsync}) that
emission in Stokes $V_{\rm sync}$
is proportional to the synchrotron intensity, $I_{\rm
sync}$ and also depends on the ratio $B_{\perp}/B_{\parallel}$, where
$B_{\perp}$ is the sky projected magnetic field and $B_{\parallel}$ is the
line-of-sight magnetic field. Following a magnetic field configuration
that is symmetric around the galactic disc, a high correlation between I
and V along the disc is expected. The level of Stokes V agrees well with
analytic calculations using simple equations for the coefficients of
the transfer equation in Appendix D. of \cite{beckert}.
\item Along a given line of sight, CP of the GSE falls off as a power
law with increasing frequency.
\item The ratio of V/I along a given line of sight, decreases with increasing frequency (See
Eq.~(\ref{eq:vsync})).
\end{itemize}
%

\section{Angular power spectra of CP due to the GSE}
\subsection{Construction of the galactic mask}
\label{sec:mask}
From Fig.~\ref{fig:hammu_map} it is clear that the galactic disc is
the highest source of foreground emission in V. Therefore, we create a
mask to block these parts of the sky in order to evaluate S/N in
Sec.~\ref{sec:detection}. The mask used in this paper 
is a superposition of a WMAP K-band mask \citep{wmap}
and a symmetric 20$\degree$ cut around the galactic plane. The galactic
plane was cut out to avoid the highest source of foreground in V. The
WMAP mask was used to remove additional point sources and generally high
synchrotron source, since it is expected 
from Eq.~(\ref{eq:vsync}) that $\eta_V$ increases
with $I_{\rm sync}$. The effective sky fraction, $f_{\rm sky}$ using
this particular mask is 0.65. 
\subsection{Angular power spectrum of galactic CP from $\hammu$}
\label{sec:power-cp}
\begin{figure}
\begin{tabular}{cc}
\includegraphics[height=0.35\textwidth]{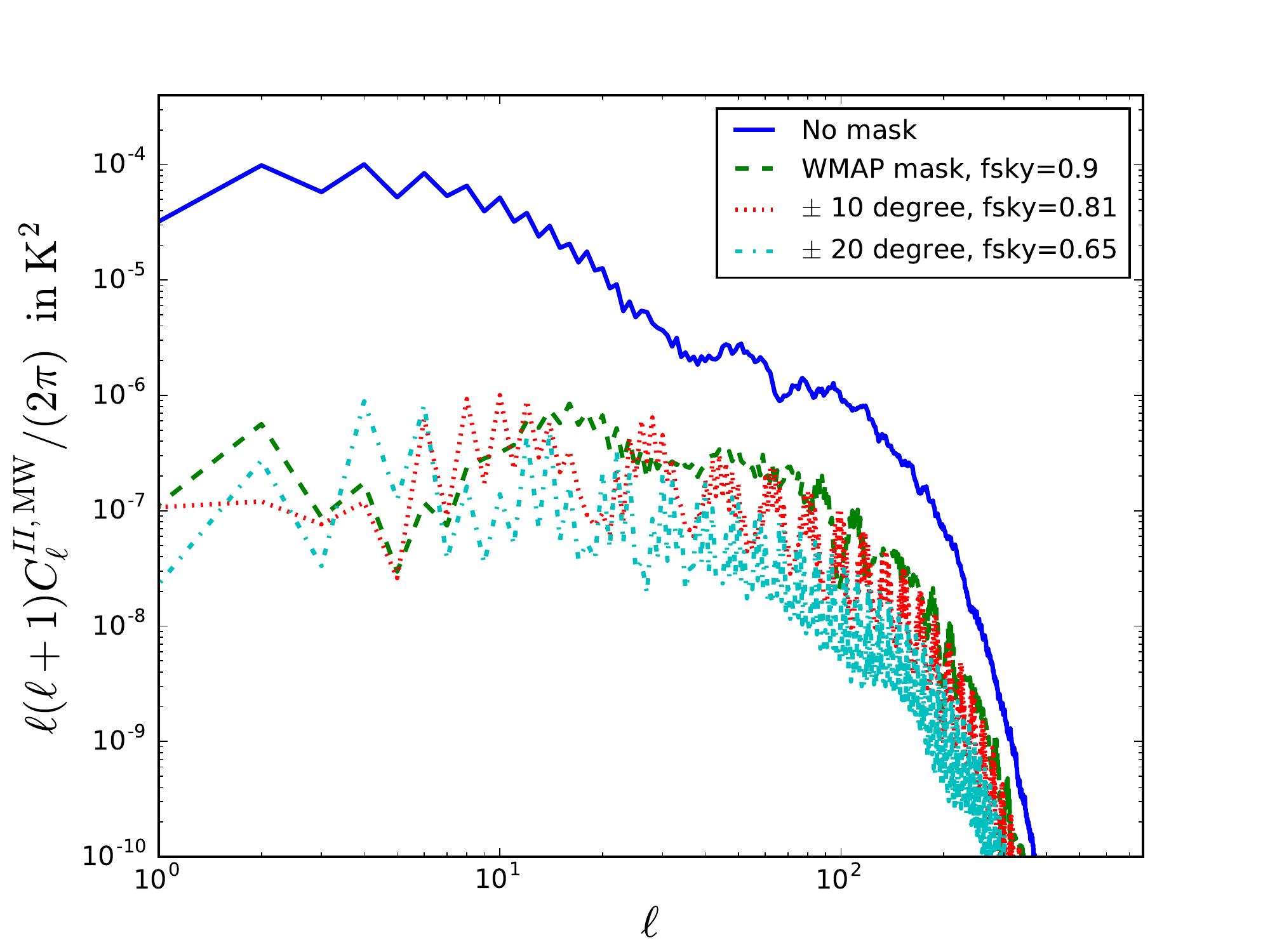} &\includegraphics[height=0.35\textwidth]{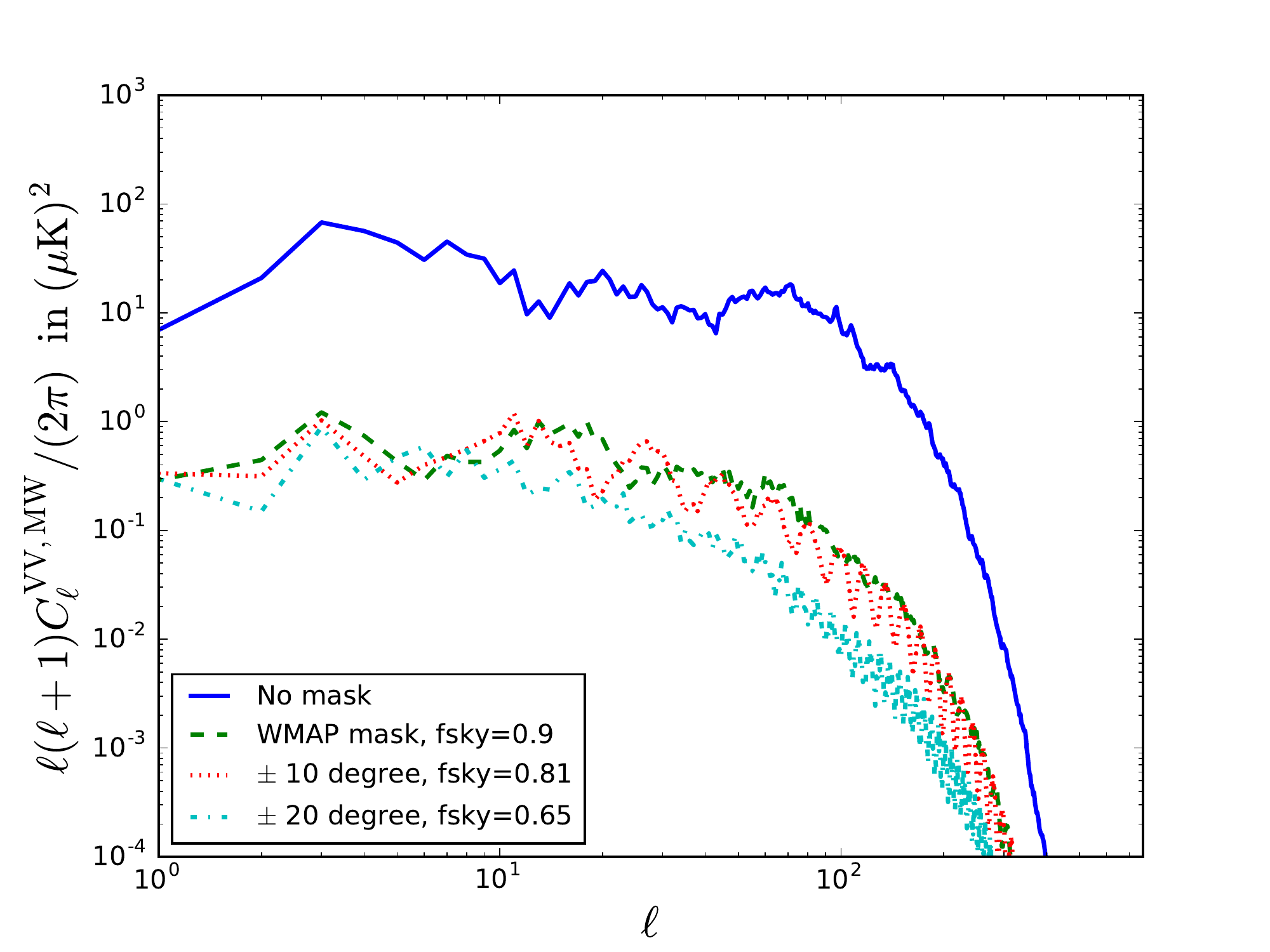} \\
\end{tabular}
\caption{
Angular power spectrum of the unpolarized synchrotron intensity in terms
of Stokes I (left
column) and 
its intrinsic Stokes V emission (right column) at $\nu$=10 GHz. Pseudo-cl values,
following Eq.~(\ref{eq:clgeneral}) and Eq.~(\ref{eq:clii}) are
plotted using different masks. Each map was smoothed using a Gaussian
beam of FWHM=1$\degree$. $\clvv$ and $\clii$ both have saw-tooth behavior
in large scale and smooth nature due to turbulence in smaller scales.
$C_{\ell}$ for unpolarized synchrotron intensity has peaks for even
values of $\ell$ and $\clvv$ has peaks for odd values of $\ell$.
Saw-tooth behavior of the power spectra simply follows from a
near-symmetric northern and southern galactic sky with very weak
dependence on longitude.}
\label{fig:clvv}
\end{figure}
In this section we will discuss the angular power spectrum of CP due to
the synchrotron
emission of the MW galaxy. The only source of CP is the intrinsic
emission term $\eta_V$ described in Eq.~(\ref{eq:transfer}).
In order to construct the
angular power spectra, we first generate the galactic circular
polarization field V=V$_{\rm sync}$($\Omega$) using $\hammu$, as a function of solid
angle $\Omega$. To calculate the power spectra, $\clvvg$ of the CP due to
the GSE, we use the following equation.
\begin{equation}
\clvvg=\frac{1}{f_{\rm
sky}(2\ell+1)}\sum_{m=-\ell}^{m=\ell}\tilde{a}_{\ell m}^*\tilde{a}_{\ell
m}
\label{eq:clgeneral}
\end{equation}
where f$_{\rm sky}$ is the sky fraction. Please see
Eq.~(\ref{eq:almtilde}) for the definition of $\tilde{a}_{lm}$. 
Angular power spectra of the
Stokes I and V components due to the GSE
are shown in Fig.~\ref{fig:clvv}.
We compute $\clvv$ using different masks, each labeled in the plot. 
Each mask is a
superposition of both WMAP K-band mask and a symmetric galactic cut. 
The synchrotron I and V fields were smoothed with a Gaussian beam of
FWHM=1$\degree$. We
make the following observations in the power spectra of CP and
unpolarized intensity, due to the GSE.
\begin{itemize}[label={--}]
\item  A saw-tooth feature in the power spectra of both Stokes I and V at
larger scale, which is due to symmetry between the northern and southern
galactic sky and invariance with shift in longitude \citep{sarkar}. A
perfectly symmetric map around the galactic plane, 
will only have angular power with non-zero
$C_{\ell}$ if only $\ell$ is even. However, if the symmetry is partially
broken, or not perfect, then $C_{\ell}$ for odd values of $\ell$
will also get populated. This is supported by an 
increasingly saw-tooth nature of the spectra in both I and V as we
remove the galactic plane using a mask and tend to a smoother and more
symmetric synchrotron
sky. 
\item The saw tooth behavior of $\clvv$ in large scale is opposite of
$C_{\ell}$, due to the extra factor of $\sin\theta$ in evaluation of
$\eta_V$ following Eq.~(\ref{eq:vsync}). This implies for a perfectly
symmetric sky in $I_{\rm sync}$, CP sky will be perfectly anti-symmetric.
This results in $\clvv$ to be non-zero for only odd values of
$\ell$. In case of a broken or imperfect anti-symmetry, $\clvv$ for even
values of $\ell$ also get populated.

\item In smaller scale, the power primarily come from turbulence and
$C_{\ell}$ for both I and CP are relatively smooth. $C_{\ell}$ falls off as $\sim$ $\ell^{-11/3}$ beyond
$\ell$ $\sim$ 50 \citep{sridhar}. The scale dependence of $\clvv$ is
also similar to $C_{\ell}$ of unpolarized synchrotron because $\eta_V$
is proportional to $I_{\rm sync}$. In the observed spectrum of
synchrotron emission at 408 MHz \citep{haslam} from the MW galaxy,
there is more power in small scales which are mostly contributed by
point sources. Since we have removed the point sources, power in small
scale drops \citep{sarkar}. 
\end{itemize}
%
\section{Detection Prospects}
\label{sec:detection}
In this section, we attempt to forecast detection prospects of a
cosmological signal of Stokes V in the CMB via current and future CMB
experiments. Let, $\clvvs$ be the signal of interest, composed of
signal from the Pop III stars, galaxy clusters 
or primordial sources. Therefore,
\begin{equation}
\clvvs=\clvvpop+\clvvgc+\clvvprim
\label{eq:signal}
\end{equation}
where $\clvvpop$ is by far the highest signal among the scenarios that are
reviewed in this paper. See Fig.~\ref{fig:signal_all} for a quick
comparison. $\clvvs$ is then dominated by the contribution
from $\clvvpop$. Detection prospects of the cosmological signal of
interest can then be simply evaluated as
\begin{equation}
\left(\frac{S}{N} \right)^2=\sum_{\ell=2}^{L_{\rm
max}}\frac{(2\ell+1)}{2}f_{\rm
sky}\left(\frac{\clvvs}{\tilde{C}_{\ell}^{\rm VV}}\right)^2
\label{eq:snr}
\end{equation}
where 
\begin{equation}
\tilde{C}_{\ell}^{VV}=\clvvs+\clvvg+\clvveg+\clnoise
\end{equation}

In Eq.~(\ref{eq:snr}), we use Eq.~(7) of \cite{decp} to evaluate
$\clvvpop$. The galactic foreground 
contribution, $\clvvg$ is primarily due to intrinsically circularly
polarized GSE. The calculation for $\clvvg$ is described in
Sec.~\ref{sec:power-cp} and
Sec.~\ref{sec:foregroundsa}-\ref{sec:foregroundsb}. The galactic signal may also include the mesospheric oxygen signal
described in Sec.~\ref{sec:foregroundsc}. 
However, we have ignored the mesospheric oxygen signal (of
CP) due to its limitation to only the largest scales. We have also set
angular power, $\clvveg$, coming from the extragalactic 
sources (that are not included in
the signal of interest in Eq.~(\ref{eq:snr})) to be zero.  
Angular power related to instrumental noise, $\clnoise$ is given by the following.
\begin{equation}
I_{\ell}^{\rm VV}=A_{P}^2exp\left(\ell^2\frac{\Theta^2_{\rm
FWHM}}{8ln2}\right)
\label{eq:noise}
\end{equation}
where,
$A_{\rm P}=\Delta_{\rm P}$(in $\mu$K/K)$\Theta_{\rm FWHM}$(in radian)T$_{\rm
CMB}$.
An important quantity, the resolution-per-pixel, $\Delta_r$ is defined
using $\Delta_{\rm P}$ given in Eq.~(\ref{eq:noise}) such that
$\Delta_r$=$\Delta_{\rm P}T_{\rm CMB}$. 
Full width at half maximum of the Gaussian beam is denoted by
$\Theta_{\rm FWHM}$.
Resolution-per-pixel is related to the detector noise-equivalent
temperature, $s$ and
total observation time, $t_{\rm obs}$ in the
following manner.
\begin{equation}
\left (A_P\right)^2=\frac{4\pi f_{\rm sky} s^2}{t_{obs}}
\label{eq:tobs}
\end{equation}
where $s$ is in the units of
$\mu$K(sec)$^{1/2}$ and $t_{\rm obs}$ is in seconds. The
Eq.~(\ref{eq:tobs}) follows from the following. The area covered by each
pixel is $\sim$ $\Theta_{\rm FWHM}^2$. Time required to get a
resolution-per-pixel of $\Delta_r$ with a detector noise-equivalent
temperature of $s$ is $\sim$
$(s/\Delta_r)^2$. Therefore, within a given observing time of $t_{\rm
obs}$, number of pixels covered will be 
N$_{\rm pix}$=$t_{\rm obs}(\Delta_r/s)^2$. Therefore, the fraction of sky area covered
by the pixels is $f_{\rm sky}=N_{\rm pix}\theta_{\rm FWHM}^2/(4\pi)$.

Goal spatial resolution depends on the type of the telescope used.
Generally, $\Theta_{\rm FWHM}$ in
arc min units is given by, 
\begin{equation}
\Theta_{\rm FWHM}=\frac{1800}{D(\rm m)\nu(\rm GHz)}
\label{eq:fwhm}
\end{equation}
where D(m) is the diameter of the telescope in meters and $\nu$ is the frequency
of CMB observation in GHz. One can therefore use a 10m telescope at 10
GHz to obtain a resolution of $\sim$ 18 arcmin. It is easier to find
dedicated observing time in smaller telescopes than the larger ones.

\begin{figure}
\begin{tabular}{cc}
\includegraphics[height=0.35\textwidth]{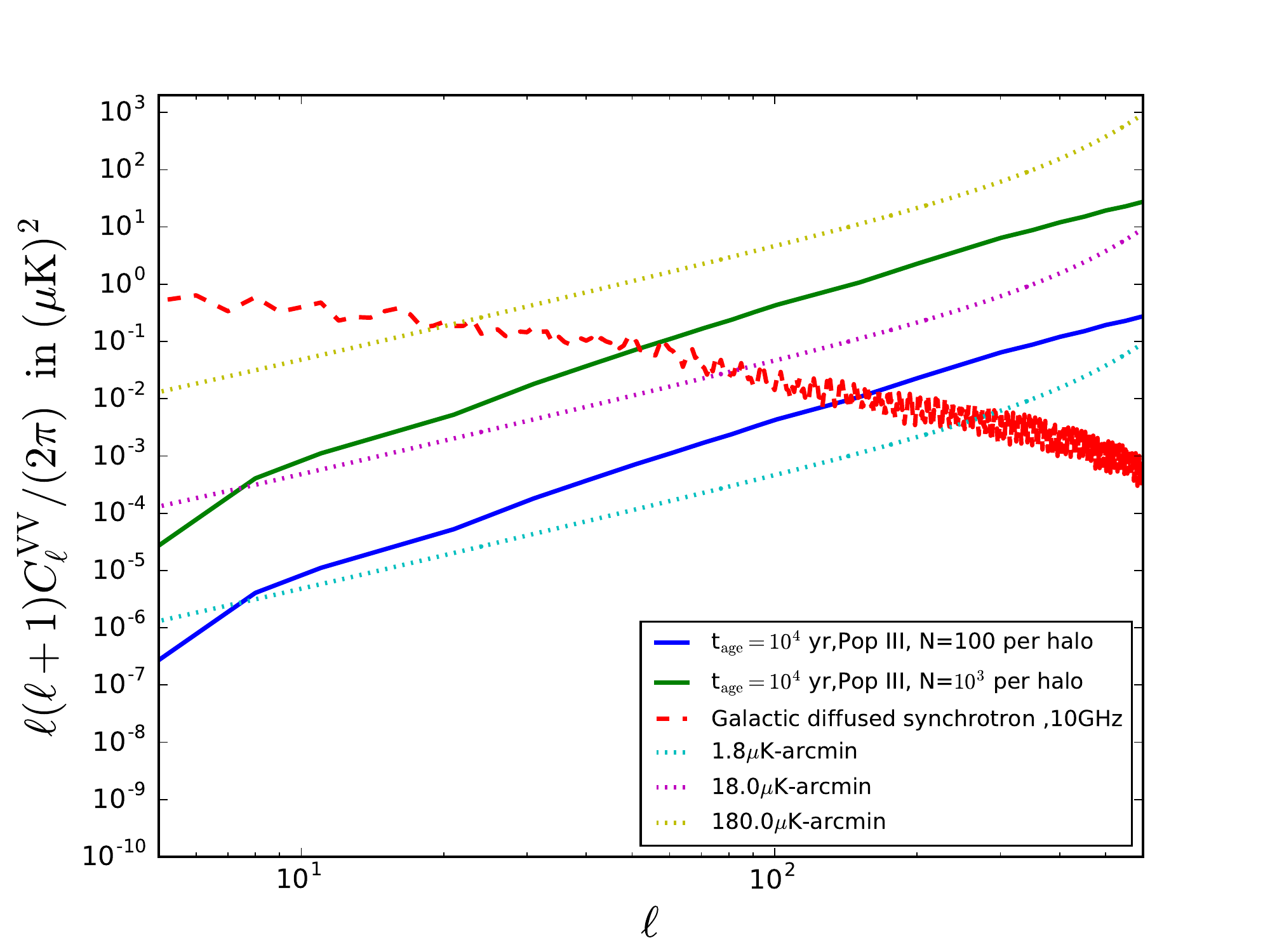}
&\includegraphics[height=0.35\textwidth]{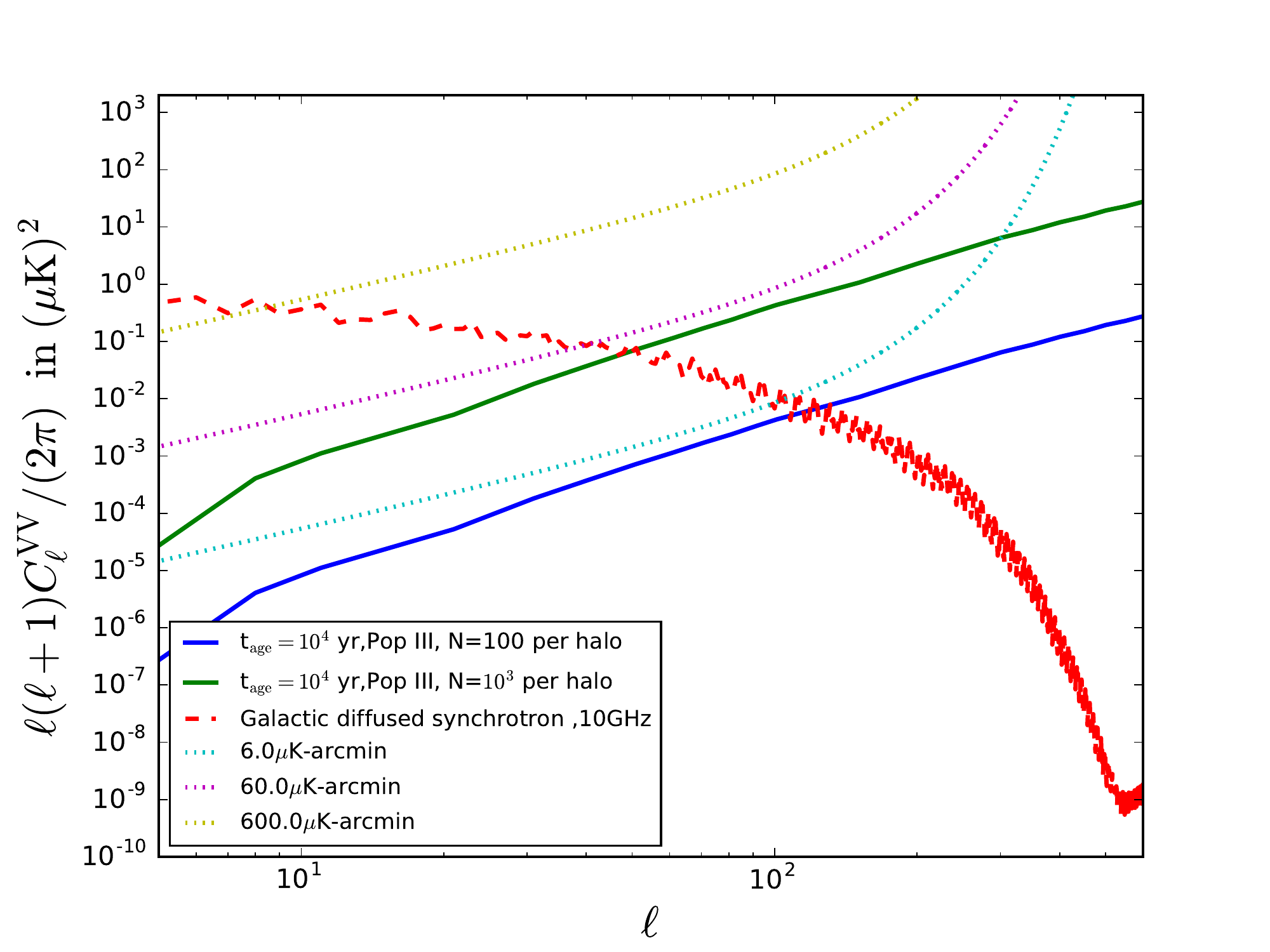} \\
\end{tabular}
\caption{
Comparison between CP signal from the Pop III stars at
$\nu$=10 GHz with FWHM=18'(left
panel), and FWHM=1$\degree$ (right panel). Noise power, plotted in
dotted lines, are calculated using
Eq.~(\ref{eq:noise}) for different values of $\Delta_{\rm P}$(in
$\mu$K/K)$\Theta_{\rm FWHM}$ (in arcmin).
Synchrotron maps were generated using Nside=256 and then smoothed using
appropriate Gaussian beam sizes. A mask of $\pm$ 20$\degree$ about the equator was used for the foreground and its angular power was evaluated
using Eq.~(\ref{eq:clgeneral}). This plot suggests that the
detectability of the cosmological CP will be limited by the detector thermal noise in small scales and by
the galactic CP in large scales. 
}
\label{fig:snr}
\end{figure}

In Fig.~\ref{fig:snr} we plot different competing factors, such as the signal,
noise and foregrounds for two different beam resolutions, 18 arcmin
(left panel) and 1$\degree$ (right panel). Resolution-per-pixel in each
case is considered to be at three different values, 0.1, 1 and 10 in the units of $\mu$K.

\begin{figure}
\begin{tabular}{c}
\includegraphics[height=0.45\textwidth]{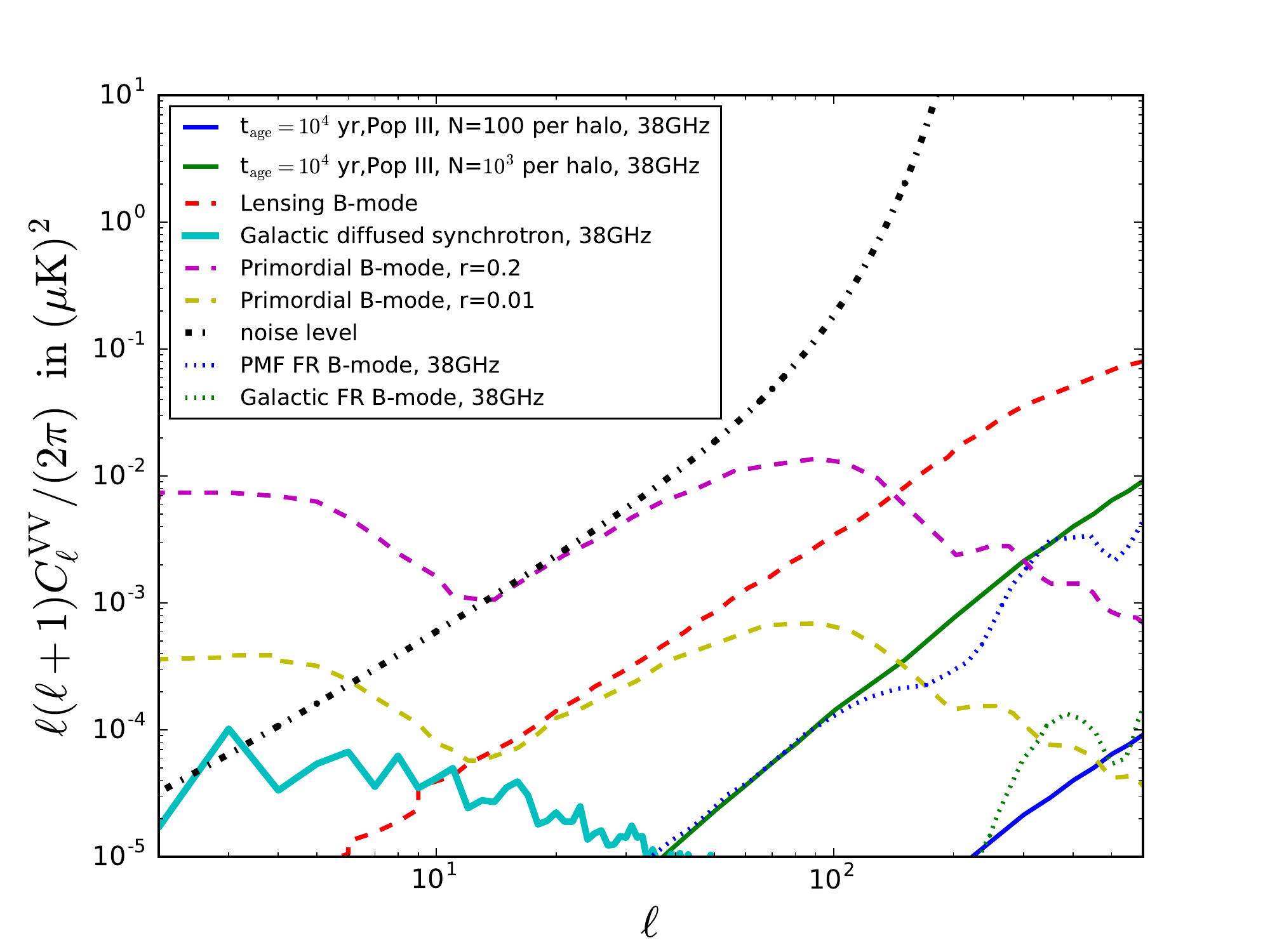}
\end{tabular}
\caption{Implications for the CLASS experiment: Angular power in CP due
the FC mechanism in the Pop III stars and intrinsic CP of the GSE are
plotted with solid lines at $\nu$=40 GHz. Frequency independent B-modes
due to gravitational lensing
and primordial gravitational waves at two different values of tensor-to-scalar ratio, $r$, are
plotted in dashed lines. The frequency dependent B-modes due to
the FR effects in the galaxy and PMF are plotted in dotted lines. Both
FR related B-mode results were taken from \cite{defr} and uses an
effective PMF strength of 1 nG today. Noise level for a survey depth of
20$\mu$K-arcmin, $\Theta_{\rm FWHM}=$1.5$\degree$ and f$_{\rm sky}$=0.5
was used to compute the noise level at 40 GHz \cite{class}. CP is used as systematics
rejection channel in experiments like CLASS and PIPER. This result
suggests that the galactic CP is important effect for CLASS at 40 GHz.
However, angular power in the galactic CP will be 
down by a factor of $(40/220)^{6.8}$$\sim$ 10$^{-5}$ at $\nu$=220
GHz (one of the frequency channels for the PIPER experiment). The
galactic CP power spectrum is derived using a $\pm$ 20$\degree$ cut
about the equator and WMAP mask to remove point sources. 
}
\label{fig:cl_class}
\end{figure}

We plot comparison between both frequency dependent (FR related) and frequency
independent (lensing and primordial gravitational waves) B-modes along
with FC generated CP and CP due to the GSE in Fig.~\ref{fig:cl_class}.
In this case the results were presented at $\nu$=40 GHz due to its
immediate relevance to the CLASS telescope. Please note that the angular
power in galactic CP will be down by a factor of (40/220)$^{6.8}$ $\sim$ 10$^{-5}$ at the frequency (220 GHz) relevant to the PIPER
telescope. The angular power in cosmological CP due to the Pop III stars
will be down by a factor of (40/220)$^6$ $\sim$ 5$\times$10$^{-5}$ at
220 GHz. Its important to note that, the galactic CP will be an
important factor to consider while probing B-modes for lower values of
tensor-to-scalar ratio, $r$. In probing B-modes, galactic CP 
is a serious effect to consider over large scales
while in smaller scales other effects such as the FR due to the galaxy
and cosmological CP could be more important. However, detectability in
smaller scales is limited by thermal noise of the detector for both
cosmological CP and B-modes.

\begin{figure}
\begin{tabular}{cc}
\includegraphics[height=0.35\textwidth]{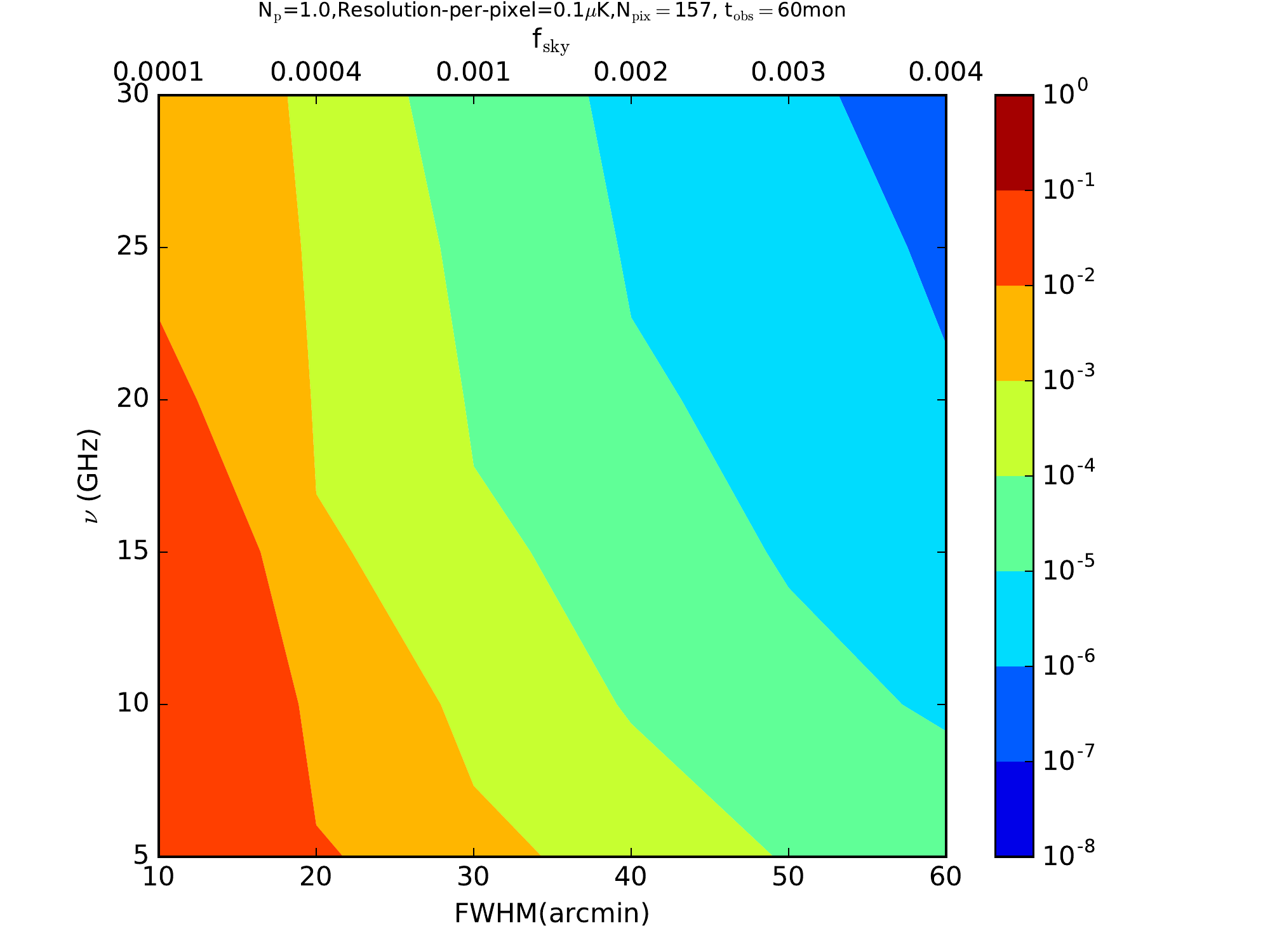}
&
\includegraphics[height=0.35\textwidth]{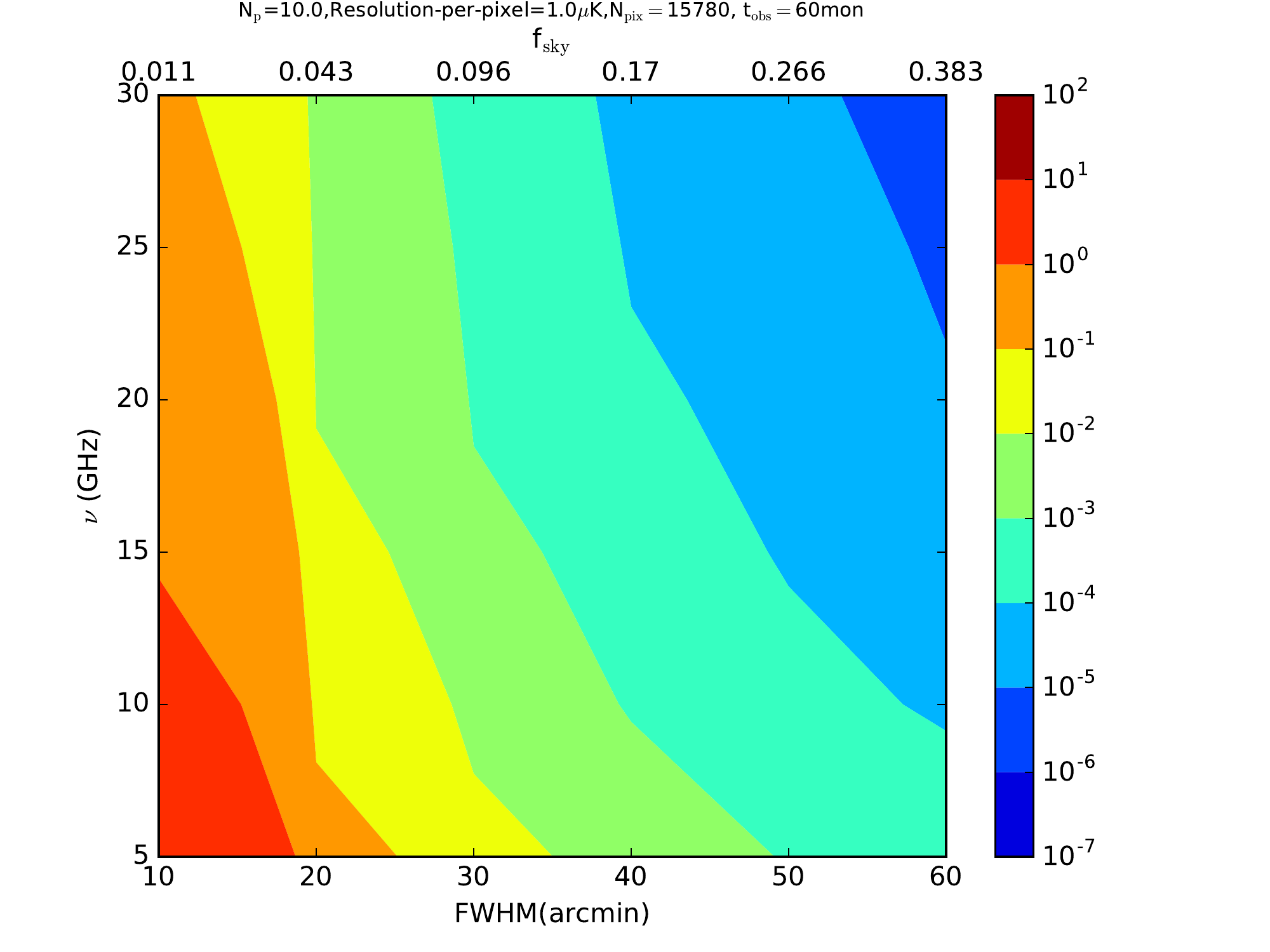} \\
\includegraphics[height=0.35\textwidth]{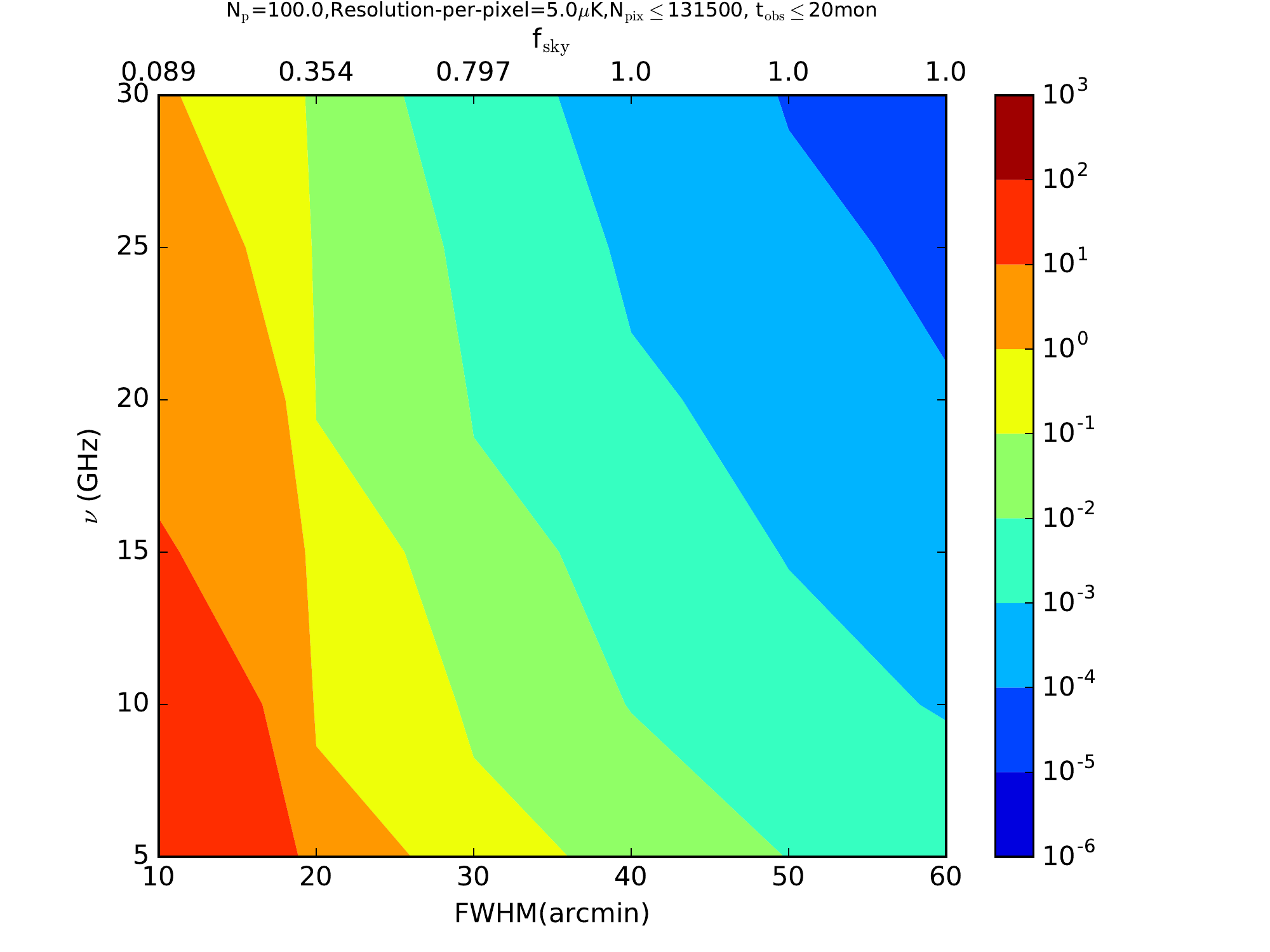}
&
\includegraphics[height=0.35\textwidth]{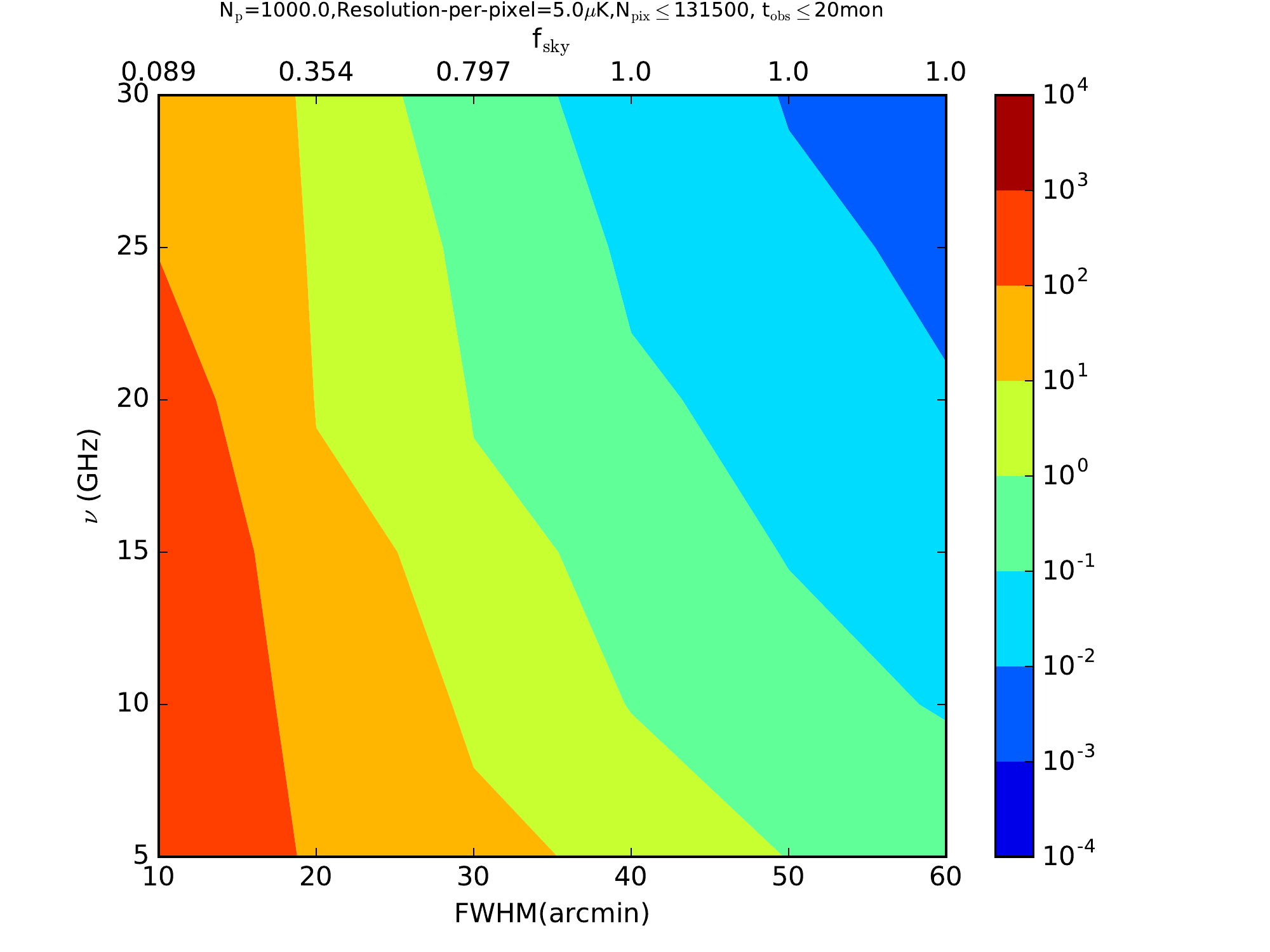} \\
\end{tabular}
\caption{S/N estimates for Pop III induced CP signal in the CMB, as a
function of CMB observation frequency, $\nu$ and resolution (FWHM) of the
Gaussian beam in arcmin.
Eq.~(\ref{eq:snr}) was
used to compute S/N at different values for the number, N$_{\rm p}$ of Pop III stars
per halo. See Sec.~\ref{sec:pop3} for more details.
Sky fractions, f$_{\rm sky}$ vary across each subplot and shown along the top margin of
each subplot. f$_{\rm sky}$ were determined from
Eq.~(\ref{eq:tobs}) for a given resolution-per-pixel (kept fixed for
each subplot) yielding an observation time, t$_{\rm obs}$=20 months or
less. A
full sky was covered in less than 20 months in the cases of high beam
size (FWHM). The
detector noise-equivalent temperature, $s$ was set at $s$=100$\mu$Ks$^{0.5}$. Note
that the signal detectability is significantly larger than unity for
N$_{\rm p}$ $\geq$ 100.
}
\label{fig:snr_grid1}
\end{figure}

\begin{figure}
\begin{tabular}{cc}
\includegraphics[height=0.35\textwidth]{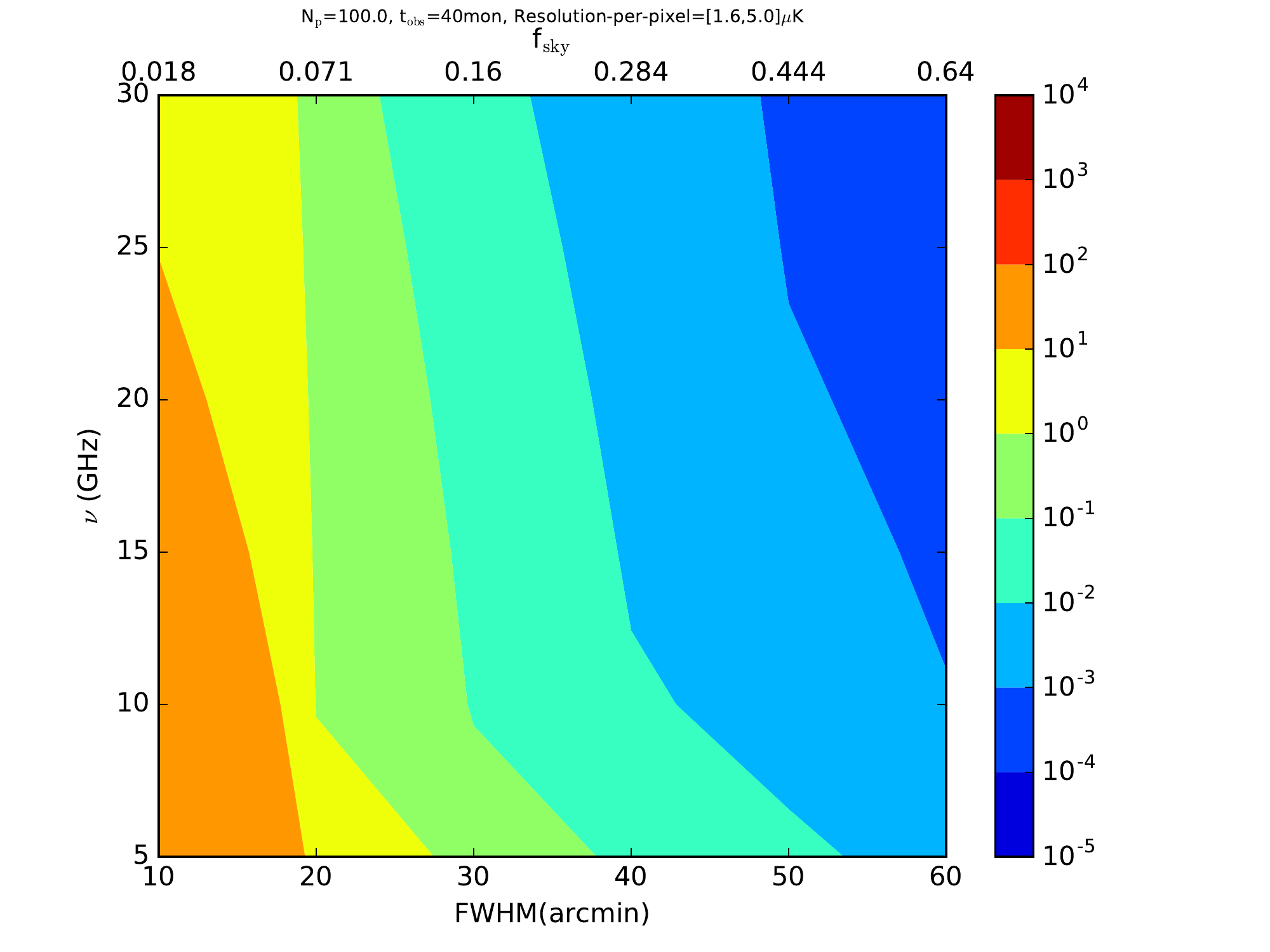}
&
\includegraphics[height=0.35\textwidth]{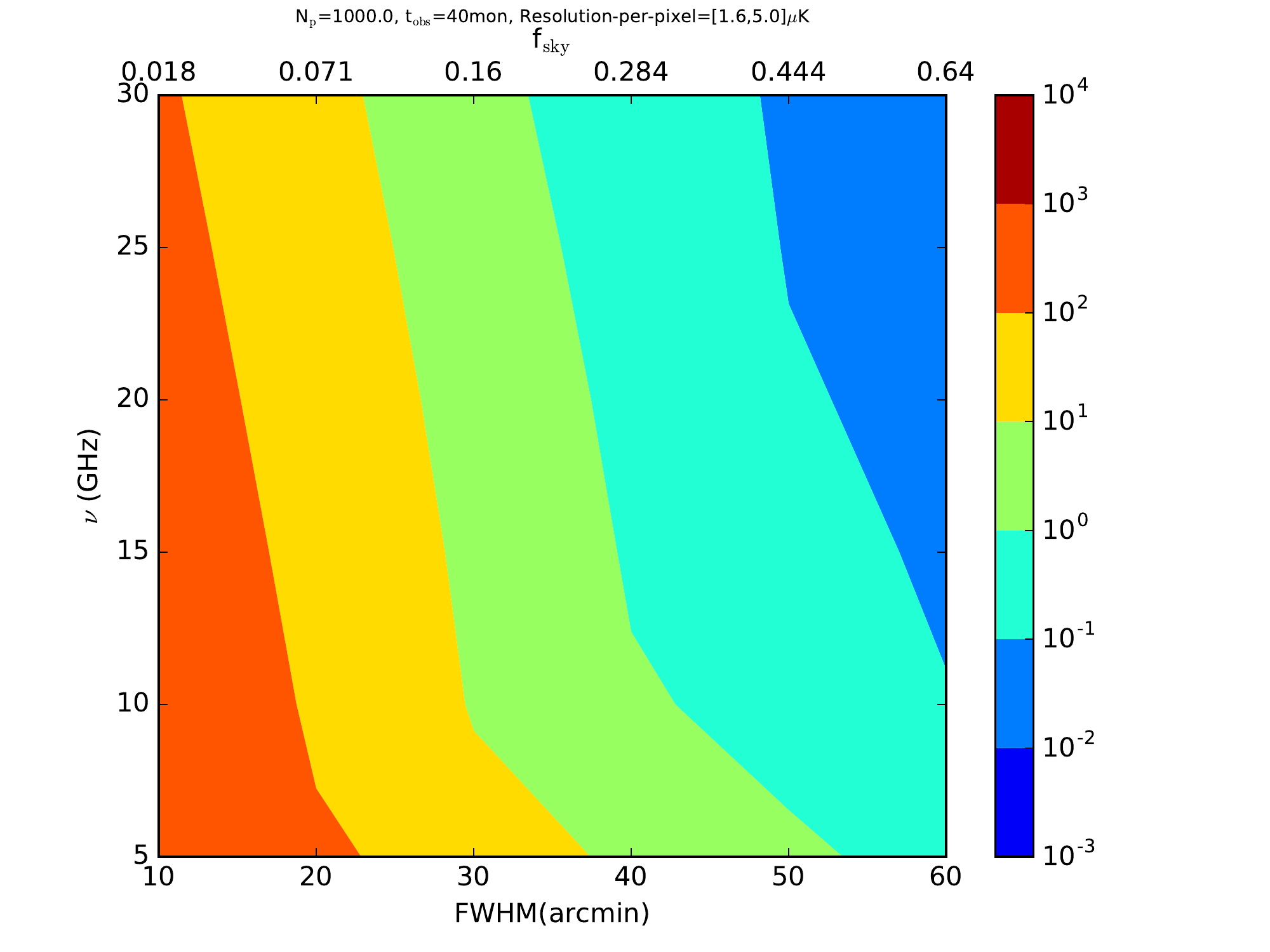}
\\ 
\end{tabular}
\caption{S/N estimates for Pop III induced CP signal in the CMB
 for different
values of the number of Pop III stars per halo, $N_{\rm p}$, as a
function of CMB observation frequency, $\nu$ and resolution (FWHM) of the
Gaussian beam in arcmin. Eq.~(\ref{eq:snr}) was
used to compute S/N. Observation time, $t_{\rm obs}$ of 40 months
was set. A sky fraction of up to 0.64 was scanned with resolution-per-pixel
determined from Eq.~(\ref{eq:tobs}). The highest value of $f_{\rm sky}$
was set at 0.64 instead of unity due to the area of the galaxy cut out
by the mask (see Sec.~\ref{sec:mask}).
The detector noise-equivalent temperature, $s$=100$\mu$Ks$^{0.5}$.
Please note that the S/N significantly larger than unity can be achieved
in 40 months of observation time, if $N_{\rm p}$ is 100 or more. Also
note that the resolution-per-pixel is varied (while keeping t$_{\rm
obs}$ fixed) in this case, in contrast
with Fig.~\ref{fig:snr_grid1} where resolution-per-pixel was fixed
with a varying t$_{\rm obs}$.}
\label{fig:snr_grid2}
\end{figure}

In Fig.~\ref{fig:snr_grid1} and Fig.~\ref{fig:snr_grid2} we present S/N
estimates for the detection of the Pop III stars as frequency and beam
resolution are varied. We consider
different values of number of Pop III stars per halo as N$_{\rm
p}$=1-1000. The main observations from the S/N estimates are the
following.

\begin{itemize}[label={--}]
\item S/N is significantly higher than unity for N$_{\rm P}$ $\geq$
100.
\item S/N increases with decreasing frequency and beam width.
\item If N$_{\rm p}$=1, S/N of higher than unity is generally not
expected at any beam resolution within the frequency range of 5-30 GHz.
\item If N$_{\rm p}$ $\geq$ 10, a S/N higher than unity is expected with
an appropriate choice of frequency and beam resolution. 
\item If N$_{\rm p}$ $\geq$ 100, a S/N significantly higher than unity
is expected with an observing time of 20 months or less. In this case,
the choice of 
frequency and FWHM of the beam is more relaxed. For example, if
N$_{\rm p}$ =1000, a S/N higher than unity can be achieved with FWHM up
to 40' and frequency of up to 50 GHz.
\end{itemize}

Some typical scenarios for observing CP involve the following.
If N$_{\rm p}$ $\geq$ 100, a S/N significantly higher than unity
is achievable using a 10m telescope at 10 GHz at 40 months of observing
time. If N$_{\rm p}$ $\geq$ 10, a S/N higher than unity
is achievable using a 10m telescope at $\sim$ 10 GHz at 60 months of observing time.

Note, that the signal of interest in our case is composed of the primordial,
Pop III star related and galaxy cluster related CP signals. However, Pop
III related CP signal is much higher (See Fig.~\ref{fig:signal_all})
and dominates other sources of CP signals of interest. For example,
S/N for solely observing the CP signal from the galaxy clusters is much
less than unity with the most optimal beam resolution and low frequency.
Therefore, a S/N higher than unity will most certainly imply the presence of CP signal
induced by the Pop III stars. 

\begin{figure}
\begin{tabular}{cc}
\includegraphics[height=0.35\textwidth]{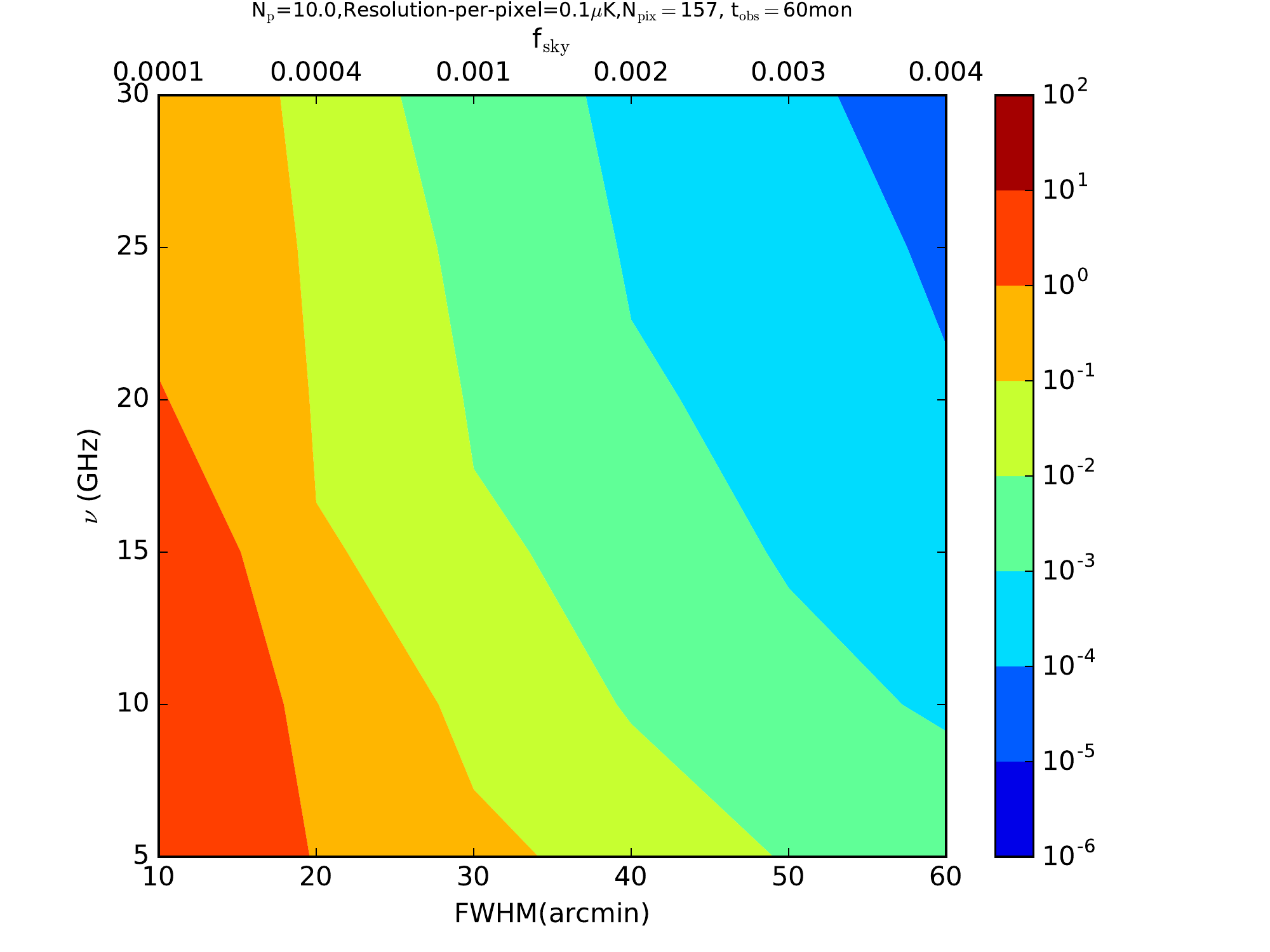}
&
\includegraphics[height=0.35\textwidth]{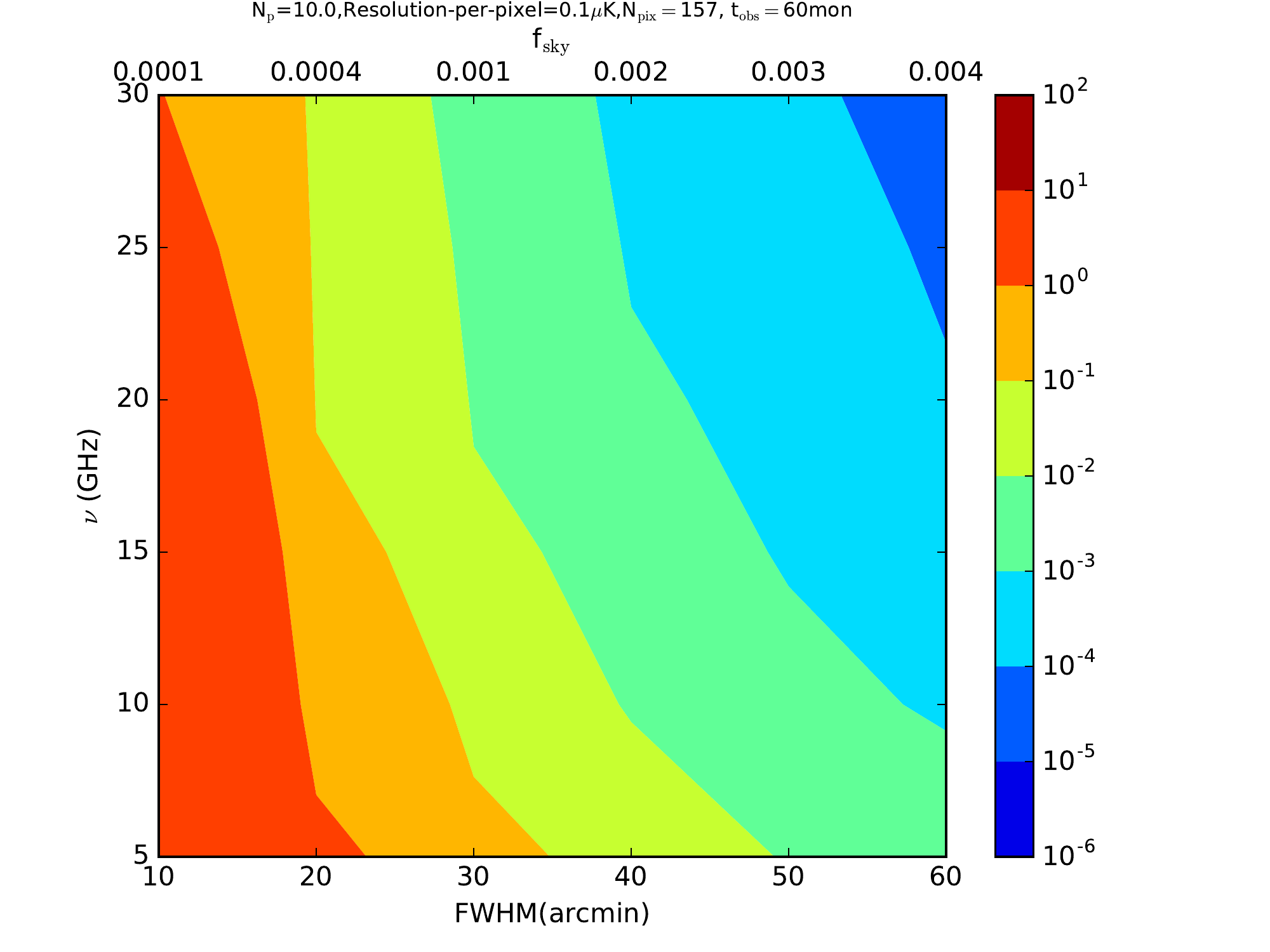} \\
\end{tabular}
\caption{S/N estimates similar to Fig.~\ref{fig:snr_grid1} with
an added possibility of partial removal of the galactic foregrounds in CP.
In this case S/N is calculated using Eq.~(\ref{eq:snr}) such that $\clvvg$
$\rightarrow$ $f_{\rm DG}\clvvg$, where $f_{\rm DG}$ $\le$ 1. Left panel
indicates $f_{\rm DG}=1.0$, assuming no removal of the galactic
foreground effects. The right panel uses $f_{\rm
DG}$=0.01, assuming a 99\% removal of the galactic effects. N$_{\rm
p}$=10 and resolution-per-pixel=0.1$\mu$K. Note that, the partial removal of
the galactic effects opens up observability in higher frequencies.
Finally, the highest limitation towards detectability come from thermal
noise of the detector.
}
\label{fig:snr_grid1_dg}
\end{figure}

In Fig.~\ref{fig:snr_grid1_dg} improvements in S/N due to partial
removal of the galaxy is considered. Partial removal of the galaxy is
achieved using a factor, $f_{\rm DG}$, using which angular power due to
galactic CP in Eq.~(\ref{eq:snr}) is modified as $\clvvg$ $\rightarrow$
$f_{\rm DG}\clvvg$. Partial removal of the galactic effects extends the
detectability prospects to higher frequencies, especially for lower
values of $N_{\rm p}$. Finally, the detectability remain limited by the
thermal noise of the detector.

\section{Discussion}
\label{sec:discussion}
In this paper we have evaluated S/N for the detection of cosmologically
important CP signals in the CMB. Frequency range of observation was
chosen to be $\sim$ 5-30 GHz. The lower limit of $\sim$ 5 GHz in the
frequency range of interest is chosen to avoid irregularities in the
synchrotron spectra due to extra-galactic self-absorbing sources.
Additionally, treatment of CP in lower frequencies need a full solution
of the polarization transfer equation in many scenarios \citep{huang}.
An
upper limit of $\sim$ 30 GHz is chosen due to sharp fall off of the CP
signal with increasing frequency.

Our work follows from \cite{decp} who showed that Pop III stars could
induce a $strong$ CP signal in the CMB. In addition to the Pop III signal,
CP signal induced in the CMB due to the galaxy clusters and other
primordial sources, such as the primordial magnetic field or symmetry
breaking mechanisms, were also considered. 
Of all the cosmologically important sources of CP
discussed in this paper, CP due to the Pop III stars dominate with
a significantly higher signal in the frequency range of interest. A
detailed description of the mechanisms that produce CP in various sources
are discussed in Sec.~\ref{sec:overview} and
Sec.~\ref{sec:transfer}. CP in the CMB is produced by
the Pop III stars via the FC mechanism which transforms an incoming
linear polarization into circular polarization in presence of an
external magnetic field.

An important foreground to the CP observation is the galactic CP due to
the GSE, which is naturally (intrinsically) circularly polarized. We
evaluate the signal
level of the CP generated from the GSE using numerical simulations
generated by the $\hammu$ code where we have implemented the calculation
of Stokes V (component corresponding to CP) following
Eq.~(\ref{eq:vsync}).

The goal frequency of observation is in the 5-30 GHz range which is
chosen due to the reasons described earlier in this section. 
There is not much known about the mass and other properties of the Pop III stars. If the number of Pop III stars per halo is as high as
1000, then observations at higher frequency (of up to 50 GHz) could lead
to a S/N higher than unity. 

The final result of our work is summarized in
Fig.~(\ref{fig:snr_grid1}-\ref{fig:snr_grid1_dg}). A S/N 
significantly higher than unity is
achievable if N$_{\rm p}$ $\geq$ 100. Generally, a S/N higher than unity
is accessible if N$_{\rm p}$ $\geq$ 10, with an appropriately chosen
frequency of observation and beam resolution. Under the most optimistic
scenario (for example, N$_{\rm p}$=1000), a S/N higher than unity is
accessible at $\nu$=50 GHz with beam resolution of up to 40'. Under the
least optimistic scenario, when N$_{\rm p}$=1, we do not expect a S/N
higher than unity.

Limitations of our results come from the current status of the GMF
models. Our results for the galactic foreground in CP is based on
numerical simulations which are partially driven by galactic synchrotron
and low frequency CMB data. The GMF models
considered in this paper are compared with Haslam data (408 MHz map), low
frequency CMB data from Planck and WMAP satellites. However, the
resolution of these datasets are at best 40'-1$\degree$, limiting our
knowledge of the GMF at smaller scales. Maps of Stokes V 
presented in this paper
depend on the GMF models we have adopted, which need to be improved to
produce the accurate maps of the GSE which match the observations more
closely. This will also improve the Stokes V maps. Finally, there is no
observed maps of the galactic CP. An all-sky map of the observed galactic CP
will reveal the nature of the GMF in greater detail, verify the theoretical
predictions on CP from the polarization transfer equations and yield
foregrounds levels to the measurements of the cosmic CP.

A low frequency measurement of the cosmic CP in the CMB can reveal
information about the Pop III stars, early universe symmetry breaking or
new physics, galaxy clusters or even the primordial magnetic field. The
signal from the Pop III stars are particularly interesting due to their
expected 
high signal level and consequently, significantly high S/N. Observing CP in the CMB
will be an indirect probe of the Pop III stars which are highly
significant as the first formed structures since the cosmic dark ages
and seeds of reionization of the universe.
Not much is known about the Pop III stars currently. Direct observation
of these high redshift
objects (z $\geq$ 15) are generally beyond the reach of future
telescopes like the JWST. Observing the CP induced in the CMB
provides a much economical way of learning about the existence and nature 
of the Pop III stars, which can even be realized in the immediate future. With the current
status of instrumentation, observing CP of the CMB provides the highest
promise in probing the Pop III stars. If the instrumentation is
improved in the near future, new physics related signal can also be explored
which currently remains far out of the reach of the highly expensive 
modern day accelerators.

Finally, an immediate practical significance of this work is towards
the cosmic B-mode exploration. This applies to the telescopes that are now being built and propose to explore primordial 
B-modes using VPM techniques. The observing strategy relies 
on using CP as a
systematics rejections channel.
This work points out that the galactic CP
effects are important in large scales, especially for frequencies below 50 GHz.  

\section*{Acknowledgements}
We thank Hiroyuki Tashiro for important discussions and suggestions as
well for a previous collaboration which inspired the current work.
We thank Tess Jaffe for her help with the $\hammu$ code. We thank Ari
Kaplan for discussions on Planck sky models simulations used to compare
with some of the $\hammu$ results. We thank the anonymous referee for an
insightful and encouraging review.


\begin{thebibliography}{99}

\bibitem{jones}
{Jones}, T.~W., \& {Odell}, S.~L. 1977, \apj, 214, 522

\bibitem{penzias} Penzias, A.~A., \& Wilson, R.~W.\
1965, \apj, 142, 419 

\bibitem{dicke} Dicke, R.~H., Peebles, 
P.~J.~E., Roll, P.~G., \& Wilkinson, D.~T.\ 1965, \apj, 142, 414 


\bibitem{cobe} Smoot, G.~F., Bennett, 
C.~L., Kogut, A., et al.\ 1992, \apjl, 396, L1 


\bibitem{dasi}
{Kovac}, J.~M., {Leitch}, E.~M., {Pryke}, C., {Carlstrom}, J.~E., {Halverson},
  N.~W., \& {Holzapfel}, W.~L. 2002, \nat, 420, 772


\bibitem{wmap} Bennett, C.~L.,
Larson, 
D., Weiland, J.~L., et al.\ 2013, \apjs, 208, 20 


\bibitem{planckgeneral} Planck 
Collaboration, Adam, R., Ade, P.~A.~R., et al.\ 2015, arXiv:1502.01582 


\bibitem{act-temp} Sievers, J.~L.,
Hlozek, 
R.~A., Nolta, M.~R., et al.\ 2013, \jcap, 10, 060 


\bibitem{spt-temp} Reichardt, C.~L., 
Shaw, L., Zahn, O., et al.\ 2012, \apj, 755, 70 


\bibitem{bicep2-lensing} BICEP2 and Keck Array Collaborations,
Ade, 
P.~A.~R., Ahmed, Z., et al.\ 2015, \apj, 811, 126 


\bibitem{polarbear} Ade, P.~A.~R., Akiba,
Y., 
Anthony, A.~E., et al.\ 2014, Physical Review Letters, 113, 021301 


\bibitem{quiet} QUIET 
Collaboration, Araujo, D., Bischoff, C., et al.\ 2012, \apj, 760, 145 


\bibitem{spt-lensing} Hanson, D., Hoover,
S., 
Crites, A., et al.\ 2013, Physical Review Letters, 111, 141301 



\bibitem{act-lensing} van Engelen, A., 
Sherwin, B.~D., Sehgal, N., et al.\ 2015, \apj, 808, 7 


\bibitem{bond} Bond, J.~R., Jaffe,
A.~H., 
\& Knox, L.\ 1998, \prd, 57, 2117 


\bibitem{mainini}
{Mainini}, R., {Minelli}, D., {Gervasi}, M., {Boella}, G., {Sironi}, G.,
  {Ba{\'u}}, A., {Banfi}, S., {Passerini}, A., {De Lucia}, A., \& {Cavaliere},
  F. 2013, \jcap, 8, 33
%
\bibitem{lubin} Lubin, P.~M., \& Smoot, G.~F.\ 
1981, \apj, 245, 1 

\bibitem{class}
Essinger-Hileman, T., Ali, A., Amiri, M., et al.\ 2014, \procspie, 9153,
91531I 

\bibitem{piper} Lazear, J., Ade, 
P.~A.~R., Benford, D., et al.\ 2014, \procspie, 9153, 91531L 
%


\bibitem{sazonov}
{Sazonov}, V.~N. 1972, \apss, 19, 25


\bibitem{westfold}
{Legg}, M.~P.~C., \& {Westfold}, K.~C. 1968, \apj, 154, 499


\bibitem{ensslin}
{En{\ss}lin}, T.~A. 2003, \aap, 401, 499


\bibitem{thomson}
{Giovannini}, M. 2009, \prd, 80, 123013


\bibitem{compton}
{Zarei}, M., {Bavarsad}, E., {Haghighat}, M., {Mohammadi}, R., {Motie}, I., \&
  {Rezaei}, Z. 2010, \prd, 81, 084035


\bibitem{noncomm}
{Aschieri}, P., {Jur{\v c}o}, B., {Schupp}, P., \& {Wess}, J. 2003, Nuclear
  Physics B, 651, 45

\bibitem{noncomm-1}
{Schaposnik}, F.~A. 2004, ArXiv High Energy Physics - Theory e-prints


\bibitem{cnub} Mohammadi, R.\ 2013, 
arXiv:1312.2199 

\bibitem{lv}{Colladay}, D., \& {Kosteleck{\'y}}, V.~A. 1998, \prd, 58, 116002

\bibitem{bromm} Bromm, V.\ 2013, Reports on 
Progress in Physics, 76, 112901 


\bibitem{pop3-properties}
{Schaerer}, D. 2002, in Astrophysics and Space Science Library, Vol. 274, New
  Quests in Stellar Astrophysics: the Link Between Stars and Cosmology, ed.
  M.~{Ch{\'a}vez}, A.~{Bressan}, A.~{Buzzoni}, \& D.~{Mayya}, 185--188


\bibitem{decp} De, S., \& Tashiro, H.\
2015, \prd, 92, 123506 


\bibitem{brommpop3openqs} Bromm, V., \& Larson, R.~B.\ 2004,
\araa, 42, 79 

\bibitem{currentprobpop3} Whalen, D.~J.\ 2013, Acta 
Polytechnica, 53, 573 


\bibitem{jwst} Gardner, J.~P.,
Mather, 
J.~C., Clampin, M., et al.\ 2006, \ssr, 123, 485 


\bibitem{pinspop3} Whalen, D.~J., Even,
W., 
Frey, L.~H., et al.\ 2013, \apj, 777, 110 


\bibitem{hypernova} Smidt, J., Whalen,
D.~J., 
Wiggins, B.~K., et al.\ 2014, \apj, 797, 97 


\bibitem{tomsim} Abel, T., Bryan, G.~L., 
\& Norman, M.~L.\ 2002, Science, 295, 93 

\bibitem{ragesim} Johnson, J.~L.,
Whalen, 
D.~J., Even, W., et al.\ 2013, \apj, 775, 107 


\bibitem{cooray}
{Cooray}, A., {Melchiorri}, A., \& {Silk}, J. 2003, Physics Letters B, 554, 1


\bibitem{beckert}
{Beckert}, T., \& {Falcke}, H. 2002, \aap, 388, 1106


\bibitem{kosowsky}
{Kosowsky}, A. 1996, Annals of Physics, 246, 49


\bibitem{weinbergbook} Weinberg, S.\ 2008, 
Cosmology, by Steven Weinberg.~ISBN 978-0-19-852682-7.~Published by
Oxford 
University Press, Oxford, UK, 2008. 


\bibitem{defr}
{De}, S., {Pogosian}, L., \& {Vachaspati}, T. 2013, \prd, 88, 063527


\bibitem{plancktau} Planck
Collaboration, Adam, R., Aghanim, N., et al.\ 2016, arXiv:1605.03507 


\bibitem{yoshida} Bromm, V., \&
Yoshida, N.\ 2011, \araa, 49, 373 


\bibitem{dunlop} Dunlop, J.~S.,
Rogers, A.~B., McLure, R.~J., et al.\ 2013, \mnras, 432, 3520 


\bibitem{loeb} Barkana, R., \&
Loeb, A.\ 2001, \physrep, 349, 125 


\bibitem{haiman} Haiman, Z.\ 1998,
Ph.D.~Thesis, 
2246 


\bibitem{oh} Oh, S.~P., Cooray, A., 
\& Kamionkowski, M.\ 2003, \mnras, 342, L20 


\bibitem{madau} Madau, P., \& Rees, M.~J.\ 2001,
\apjl, 551, L27 

\bibitem{suwa} Suwa, Y., Takiwaki, T., 
Kotake, K., \& Sato, K.\ 2008, First Stars III, 990, 142 


\bibitem{frebel} Frebel, A., Aoki, W., 
Christlieb, N., et al.\ 2005, \nat, 434, 871 


\bibitem{tumlinson} Tumlinson, J., 
Venkatesan, A., \& Shull, J.~M.\ 2004, \apj, 612, 602 


\bibitem{greif} Greif, T.~H., Glover, 
S.~C.~O., Bromm, V., \& Klessen, R.~S.\ 2010, \apj, 716, 510 


\bibitem{wyithe} Wyithe, J.~S.~B., \& Loeb, A.\ 2003,
\apjl, 588, L69 

\bibitem{syncpop3} Meiksin, A., \& Whalen, D.~J.\
2013, \mnras, 430, 2854 


\bibitem{pop3-xu}
{Xu}, H., {Wise}, J.~H., \& {Norman}, M.~L. 2013, \apj, 773, 83


\bibitem{reviewpmf}
{Grasso}, D., \& {Rubinstein}, H.~R. 2001, \physrep, 348, 163


\bibitem{ruth} Durrer, R., \& Neronov, A.\ 2013,
\aapr, 21, 62 

\bibitem{planckpmf} Planck 
Collaboration, Ade, P.~A.~R., Aghanim, N., et al.\ 2015,
arXiv:1502.01594


\bibitem{cowsik} Cowsik, R., \& Mitteldorf, J.\
1974, \apj, 189, 51 


\bibitem{beckreview} Beck, R., \& Wielebinski, R.\
2013, Planets, Stars and Stellar Systems.~Volume 5: Galactic Structure
and Stellar Populations, 5, 641 


\bibitem{wmapy1fore} Bennett, C.~L.,
Hill, 
R.~S., Hinshaw, G., et al.\ 2003, \apjs, 148, 97 


\bibitem{wmapy7fore} Gold, B., Odegard, N., 
Weiland, J.~L., et al.\ 2011, \apjs, 192, 15 


\bibitem{crab} Andrew, B.~H., \&
Purton, C.~R.\ 1967, \nat, 215, 493 


\bibitem{crab2} Tademaru, E.\ 1973, \apj,
183, 625 


\bibitem{planckdiffuse} Planck
Collaboration, Ade, P.~A.~R., Aghanim, N., et al.\ 2014, \aap, 571, A12 



\bibitem{page} Page, L., Hinshaw, G., 
Komatsu, E., et al.\ 2007, \apjs, 170, 335 


\bibitem{strong} Strong, A.~W., 
Moskalenko, I.~V., \& Reimer, O.\ 2004, \apj, 613, 962 


\bibitem{smootsync} Smoot, G.~F.\ 1999, 
arXiv:astro-ph/9902201 


\bibitem{oxygen} Hanany, S., \& Rosenkranz, P.\
2003, \nar, 47, 1159 


\bibitem{waelkens} Waelkens, A., Jaffe, T., Reinecke, M.,
Kitaura, F.~S., \& En{\ss}lin, T.~A.\ 2009, \aap, 495, 697 


\bibitem{healpix} G{\'o}rski, K.~M., 
Hivon, E., Banday, A.~J., et al.\ 2005, \apj, 622, 759 


\bibitem{sun} Sun, X.~H., Reich, W., Waelkens, A., \&
En{\ss}lin, T.~A.\ 2008, \aap, 477, 573 


\bibitem{han} Han, J.~L., Ferriere,
K., 
\& Manchester, R.~N.\ 2004, \apj, 610, 820 


\bibitem{ne2001} Cordes, J.~M., \&
Lazio, T.~J.~W.\ 2002, arXiv:astro-ph/0207156 


\bibitem{haslam} Haslam, C.~G.~T., Salter, C.~J.,
Stoffel, H., \& Wilson, W.~E.\ 1982, \aaps, 47, 1 


\bibitem{rema} Remazeilles, M.,
Dickinson, C., Banday, A.~J., Bigot-Sazy, M.-A., \& Ghosh, T.\ 2015,
\mnras, 451, 4311 


\bibitem{psmpaper} Delabrouille, J., Betoule, M., Melin,
J.-B., et al.\ 2013, \aap, 553, A96 


\bibitem{sarkar} Mertsch, P., \& Sarkar, S.\ 2013,
\jcap, 6, 041 


\bibitem{sridhar} Goldreich, P., \& Sridhar, S.\
1997, \apj, 485, 680 


\bibitem{huang}
{Huang}, L., {Liu}, S., {Shen}, Z.-Q., {Cai}, M.~J., {Li}, H., \& {Fryer},
 C.~L. 2008, \apjl, 676, L119

%
%
%
%
%
%
%
%
%
%
%
%
%
%
%
%
%
%
%
%
%
%
%
%
%
%
%
%
%
%
%
%
%
%
%
%
%
%
%
%
%
%
%
%
%
%
%
%
%
%
%
%
%
%
%
%
%
%
%
%
%
%
%
%
%
%
%
%
%
%
%
%
%
%
%
%
%
%
%
%
%
%
%
%
%
%
%
%
%
%
%
%
%
%
%
%
%
%
%
%
%
%
%
%
%
%
%
%
%
%
%
%
%
%

\end{thebibliography}
\end{document}